# A Non-topological Extension of Bending-immune Valley Topological Edge States

Tianyuan Liu[1,2], Wei Yan[1,2] and Min Qiu[1,2]

1. Key Laboratory of 3D Micro/Nano Fabrication and Characterization of Zhejiang Province, School of Engineering, Westlake University, 18 Shilongshan Road, Hangzhou 310024, Zhejiang Province, China

2. Institute of Advanced Technology, Westlake Institute for Advanced Study, Shilongshan Road, Hangzhou 310024, Zhejiang Province, China

**Abstract:**

Breaking parity (**P**) symmetry in $C_6$ symmetric crystals is a common routine to implement a valley-topological phase. At an interface between two crystals of opposite valley phases, the so-called valley topological edge states emerge, and they have been proven useful for wave transport with robustness against 120° bending and a certain level of disorder. However, whether these attractive transport features are bound with the valley topology or due to topological-irrelevant mechanisms remains unclear. In this letter, we discuss this question by examining transport properties of photonic edge states with varied degrees of the **P**-breaking that tune the valley topology, and reveal that the edge states preserve their transport robustness insensitive to the topology even when the **P**-symmetry is recovered. Instead, a unique modal character of the edge states—with localized momentum hotspots around high-symmetric **K** (**K'**) points—is recognized to play the key role, which only concerns the existence of the valleys in the bulk band structures, and has no special requirement on the topology. The "non-topological" notion of *valley edge states* is introduced to conceptualize this modal character, leading to a coherent understanding of bending immunity in a range of edge modes implemented in $C_3$ symmetric crystals—such as valley topological edge states, topological edge states of 2D Zak phase, topological-trivial edge states and so on—, and to new designs in general rhombic lattices—with exemplified bending angle as large as 150°.

**Introduction:**

Topology, a mathematical concept of classification, has spawned a new perspective in physics in the past decades[1-4]. When two crystals with gapped bands of different topological phases are interfaced, gapless edge states emerge, which are usually robust to disorders and defects[2,5,6]. To implement non-trivial topological phases, various approaches have been proposed according to topological classifications. Taking 2D systems for example, one can break the time reversal (**T**) symmetry to generate Chern insulator[7-9] or utilize spin or valley degree of freedom (DOF) to generate **Z**$_2$ insulator under **T**-symmetry preservation[10-13].

Recently, valley-topological insulator has drawn lots of attention due to its simple implementation, particularly in photonic aspects[13]. It usually starts with a hexagonal lattice of six-fold rotational (C$_6$) symmetry (Fig.1a). Under the protection of the parity-time (**PT**) symmetry, Dirac cones exist at high symmetric **K/K'** points (Fig.1b). By detuning sublattices A, B, the **P** symmetry is broken, resulting in bandgap opening and valley-topological phase (Figs.1b and c). We illustrate this procedure with a 2$^{nd}$ order tight-binding model, with $\varepsilon_A$, $\varepsilon_B$ denoting the on-site energy of A, B sublattices, $t_1$ and $t_2$ representing the nearest (A-B) and next-nearest (A-A and B-B) hopping energies, respectively. The 2-band Hamiltonian is

$$\widehat{H} = \sum_k \psi^\dagger \begin{bmatrix} \varepsilon_A - t_2 \sum_\alpha e^{i\mathbf{k}\cdot\boldsymbol{\alpha}} & -t_1 \sum_\delta e^{i\mathbf{k}\cdot\boldsymbol{\delta}} \\ -t_1 \sum_\delta e^{-i\mathbf{k}\cdot\boldsymbol{\delta}} & \varepsilon_B - t_2 \sum_\alpha e^{i\mathbf{k}\cdot\boldsymbol{\alpha}} \end{bmatrix} \psi,$$  Eq. 1

with $\psi = [\hat{a}_k; \hat{b}_k]$ and $\psi^\dagger = [\hat{a}_k^\dagger, \hat{b}_k^\dagger]$, where $\hat{a}_k, \hat{a}_k^\dagger$ ($\hat{b}_k, \hat{b}_k^\dagger$) are the annihilation and creation operators for A (B) sublattice. $\boldsymbol{\delta}$ and $\boldsymbol{\alpha}$ are A-B and A-A/B-B hopping vectors, respectively (Fig. 1a). When A, B are detuned with a tiny detuning ratio $\Delta \equiv (\varepsilon_B - \varepsilon_A)/t_1$, Berry curvatures $\Omega_n(k)$ are localize at **K** and **K'** points (Fig.1c), the integration of which over the half Brillouin zone of the valence band gives the valley-Chern number $C_v \sim \gamma \text{sgn}(\Delta)1/2$ with $\gamma = 1$ (-1) for **K** (**K'**) valley [14]. Interfacing two bulks of opposite $C_v$ (Fig.1d), gapless edge states locked with each valley emerge (Fig.1f). Such gapless edge states are called *valley-topological edge states* and have been reported to be robust against 120° bending and a certain level of position disorder[15,16]. Utilizing them, a variety of integrated photonic components has been put forward and demonstrates excellent performances, including in topological laser[17-19], slow light waveguide[20,21], high fidelity transmission line[15,16] and quantum circuits[22,23].

Previous literatures mainly discussed valley-topological edge states under an implicit assumption that the valley-topology, attached to the localized Berry curvatures at **K** and **K'** points, is well defined. However, as the detuning ratio $\Delta$ (i.e., **P**-breaking) increases, the initially localized Berry curvatures become dispersed and crosstalk among contrasting valleys (Fig.1g). Consequently, the valley-Chern number cannot be well defined anymore, thereby turning the edge states from gapless into gapped. Particularly, when the detuning is sufficiently large that dwarfs the A-B hopping, A and B sublattices become decoupled, so the **P**-symmetry in terms of decoupled A and B subsystems are recovered. In this case, the Berry curvatures approach null, therein the edge states become topologically trivial (Fig.1h).

At this point, it is natural to quest how the transport characteristics of the edge states shall evolve as the valley-topology phases vary? Moreover, with respect to the consensus that the topology is generally favorable for the associated edge states to resist perturbations, we wonder whether the valley-topology is an essential condition for the widely reported transport robustness. Besides their theoretical meanings, these questions are relevant to and intensified by a few recent experiments that observe similar propagation properties of the edge states through sharp bending and random defects with and without the valley-topology[24,25]. So far, no satisfactory interpretation has been presented. Responsively, it becomes necessary to elucidate the role of the valley-topology in light transport that is largely regarded as an advantage. We discuss these questions below.

**Evolution of Valley-Topology and Photonic Edge States**

To inspect the evolution of valley-topology as well as transport properties of edge states during the **P**-breaking procedure, we construct a hexagonal PhC. The sublattice A (dark dots) and B (white dots) are Si rods ($n = 3.4$) in air background, see the inset in Fig.2a. The radii of the A ($r_A$) and B ($r_B$) rods are initially set to be $0.2a$ with lattice constant $a$. The **P**-breaking is implemented by reducing the radius of the rod B, which equivalently increases the resonant frequency $\varepsilon_B$ in the tight-binding model. The edge states are generated by cutting the PhC bulk along the zig-zag direction and interfacing it with a mirror counterpart that permutes A and B rods (similar to Fig.1d).

We calculate the spectra of bulk photonic modes and edge states of $E_z$-polarization with varied detuning ratio $D \equiv (r_A - r_B)/r_A \in [0,1]$. As illustrated in Fig.2a, the edge states (red curves in the band diagrams) originate from the **P**-breaking with a non-zero $D$. Similar to the insights from the tight-binding model, the edge states are gapless when $D$ is tiny where the valley-Chern number is well defined. Further increasing $D$, the edge states turn to be gapped, i.e., losing their topological character. Finally, when the sub-lattice $B$ is eliminated ($D = 1$), the asymmetric lattice becomes symmetric, possessing zero Berry curvature due to the simultaneous **P**- and **T**-symmetry. In this **P**-breaking and -recovery procedure, the edge states maintain their presence without distinct degradation in spectral bandwidth.

**Universal Transport Robustness of Topological and Topological-trivial Edge States**

We first examine the bending-immune properties of the edge states through a Z-shaped waveguide with 120° bending angle at different detuning ratios. The edge states are guided along the zig-zag interface as marked with black lines (see Fig.2b and c). The calculations reveal a counterintuitive phenomenon, the 120° bending immune property, which is widely regarded the privilege of the valley-topology, can exist for the topological-trivial edge states. As is shown in Fig.2d, during the whole detuning procedure, a passband with ~100% transmittance, coinciding with the spectra of the edge states, is always present. This suggests that the bending immune property of the edge states is independent of the detuning and thus of the valley-topology, even when the $D$=1 where the PhC bulks regain the P-symmetry and completely lose the valley-topology.

We then evaluate the robustness of the edge states against disorders, which has been argued as another advantage of the valley-topology[26]. The disorders are introduced by randomizing the positions of Si rods in a straight waveguide obeying a normal distribution (see inset of Fig.2e), of which the standard deviation σ measures disorder strength (see supplementary Sec.1.2 for simulation details). Figure 2e demonstrates the back-scattering free length (ξ) via normalized disorder ratio ($\sigma/a$) for PhCs with detuning ratio $D$=0.5 (approximately topological) and $D$=1 (absolutely trivial). First, when the edge states in both cases are set corresponding to **K**-point (square marks), the topological edge states (*D*=0.5, red) are observed to perform better than the non-topological ones (*D*=1, blue), but nevertheless of the same order of magnitude. However, in this comparison, the group indices of the two cases are unequal. For a fairer assessment, we then set their operating frequencies to realize the same group indices $n_g \simeq 100$ (round marks). In this situation, the non-topological edge states (blue) even demonstrate slightly better robustness to disorders compared with the topological edge states (red).

Therefore, we show that valley-topological trivial edge states can demonstrate qualitatively similar robustness against sharp bending and random disorder compared with topological cases, which is in concert with recent experiments[24,25].

**Momentum matching, a non-topological perspective for bending immunity**

The common transport features of the edge states in both topological and topological-trivial scenarios suggest that the underlying physical mechanism might be topology-independent. Essentially, the transport of the edge states under perturbations is dictated by wave scattering. In a bending problem, scatters are considered as the permittivity redistribution from a straight waveguide to the bending one, denoted by $\Delta\varepsilon$. The involved waves are incident, reflected and transmitted edge states (assuming single-mode guidance), whose normalized electric fields are denoted by $\psi_i$, $\psi_r$ and $\psi_t$, respectively. Under the 1$^{st}$ order perturbation approximation, the reflection and transmission measures of the edge states are given by the modal overlap terms, $A_\alpha \equiv \int d\mathbf{r}\, \psi_\alpha^*(\mathbf{r})\Delta\varepsilon(\mathbf{r})\psi_i(\mathbf{r})$ with $\alpha = \{r, t\}$. Such terms can be transformed into the *k*-space as $A_\alpha = (2\pi)^2 \int d\mathbf{k}_i \int d\mathbf{k}_\alpha\, \psi_\alpha^*(\mathbf{k}_\alpha)\psi_i(\mathbf{k}_i)\Delta\varepsilon(\mathbf{k}_\alpha - \mathbf{k}_i)$, underlining that an incident mode is scattered into an reflected or bended mode unitizing the momentum $\mathbf{k}_\alpha - \mathbf{k}_i$ supplied by the scatters.

Figure 3b plots the momentum profile of the scatters, $|\Delta\varepsilon(\mathbf{k})|$, for a typical bending waveguide sketched in Fig.3a. The scatters are approximated as points for simplicity. First, $|\Delta\varepsilon(\mathbf{k})|$ displays periodic hotspots on reciprocal lattice points, revealing crystal momentum. Moreover, attaching to the crystal-momentum points, bright lines in vertical and 150-degree oblique directions—perpendicular to the two interfaces of the scatter region (cf. Fig.3a)—extend out, which are the momentum provided by the interfaces and physically responsible for interfacial specular reflections.

We then examine the momentum profiles of the edge states within the tight-binding model (Fig.1) for physical transparence. The wavenumber of the incident mode $\psi_i$ is set equal to the projection of **K**-point onto the propagation direction (i.e., along the zig-zag interface), and the benefits of this choice shall become clear later. In the topological case with a tiny AB detuning, $|\psi_i(\mathbf{k})|$ (incidence, Fig.3d) and $|\psi_t(\mathbf{k})|$ (transmission, Fig.3e) show hotspots localized at the **K** points, while the hotspots of $|\psi_r(\mathbf{k})|$ (reflection, Fig.3f) are localized at the **K'** points. As the detuning ratio $\Delta$ increases that mitigates the valley-topology, these hotspots are still preserved, as is shown in Fig.3g-3i that corresponds to the trivial case with sublattice B eliminated. This consistence in momentum profiles is irrelevant to valley-topology, as analytically discussed in the supplementary Sec.3. Note that the momentum gap between $|\psi_i(\mathbf{k})|$ and $|\psi_r(\mathbf{k})|$ does not match with either the crystal or interface momenta of the scatters, thus resulting in a tiny $|A_r|$. In contrast, the similar profiles of $|\psi_i(\mathbf{k})|$ and $|\psi_t(\mathbf{k})|$ leads to a large value of $|A_t|$. Consequently, in both the topological and topological-trivial cases, we have $|A_t| \gg |A_r|$ (see Supplementary Sec.2 for more discussions),

implying the reflection suppression. As the wavenumber of the edge state deviates from the (projected) **K** or **K'** point but is still distant to the **Γ** and **M** points where group indices diverge, the large momentum separation between $|\psi_i(\boldsymbol{k})|$ and $|\psi_r(\boldsymbol{k})|$ still ensures $|A_t| \gg |A_r|$. These observations thus interpret why the edge states can perfectly transmit through the 120° bending insensitive to the valley-topology as observed in Fig.2. They also explain that the edge states in both topological and topological-trivial cases perform similarly against disorders because their momentum profiles are close.

Not every edge state with the wavenumber close to the projected **K** or **K'** points can support bending immunity, even in the same system. Figure 3c demonstrates the projected band diagram of the edge states for the topological-trivial case considered in Fig.3g-i. Besides the upper branch of the edge states that is spectrally close to the bulk valleys, there additionally exists an unexamined lower branch (red). Unlike the upper branch, the edge state in this branch, when the wavenumber is set according to the projection of the **K** point, shows hotspots localized at the projection of the **K** point to the $\boldsymbol{\Gamma} - \boldsymbol{K}'$ intercept, as shown in Fig.3k-l. In this case, the modal profiles of $|\psi_i(\boldsymbol{k})|$ and $|\psi_t(\boldsymbol{k})|$ do not coincide, so $|A_t| \gg |A_r|$ is no longer valid and the bending transmittance is severely diminished (see the right panel of Fig.3c). In addition, if we interface two PhCs along an armchair direction, due to the valley mix[27], the momentum separation between $|\psi_i(\boldsymbol{k})|$ and $|\psi_r(\boldsymbol{k})|$ will be decreased and the bending immunity is also unattainable (supplementary Sec.5).

The above momentum-matching analyses suggest that it is not the valley-topology but the unique modal profile of the edge states—with momentum hotspots at the **K/K'** points—that plays a key role in suppressing bending reflection. More specifically, this modal character renders that the incident and bended states have a perfect momentum matching—recalling that the **K/K'** points are invariant under the rotation of 120° bending angle—, while the momentum separation between the incident (**K/K'**) and reflection states (**K'/K**) cannot be efficiently remedied by the scatter momenta. In this way, the bending immunity is enabled.

**Valley edge states (VESs), a class of edge states that supports bending immunity**

To conceptually unify the bending-immune edge states observed in topological and topological-trivial cases, they are collectively referred as *valley edge states (VESs)*: *A branch of edge modes supported by two interfaced hexagonal PhCs can be regarded as VESs if they, at specific wavenumbers equal to the projected* **K** *or* **K'** *points, have their momentum hotspots localized around* **K** *or* **K'** *points, similar to bulk Bloch modes at* **K** *or* **K'** *points*. The meaning and importance of "*valley*" shall be clarified later. According to this definition, the edge states studied in Fig.2 and also the upper branch of the edge states in Fig.3c are the VESs, while the lower branch of the edge states in Fig.3c is not.

The stressed momentum characteristics of the VESs actually inherits from the **K/K'** valleys in the bulk band structure (due to the C$_3$ symmetry). It is known that an edge state can be decomposed into a linear combination of bulk Bloch modes, which constitute a complete basis. From the linear expansion analysis (see supplementary Sec.4), the expansion coefficients of the bulk mode scales with $1/(\omega_e - \omega_b)$ with $\omega_e$ and $\omega_b$ representing the eigen-frequency of edge and bulk modes. Since the **K/K'** valleys are frequency extrema of the bulk bands, for a neighboring edge state with wavenumber corresponding projected **K/K'** points, its modal profile should be primarily composed of the bulk modes around the valleys, thus rending its momentum hotspots distributed around the **K/K'** points. In view of this, it is interesting to note that the bending immunity of the VESs is a result of the intrinsic property of the bulk crystal, which is spiritually close to the bulk-edge correspondence in the topological physics.

**Revisiting bending immunity with VESs**

The VESs concept provides a new perspective to coherently understand bending immunity of a set of edge states implemented in $C_3$ PhCs, such as the valley-topological and valley-topological trivial edge states as discussed above (see the summarization in Fig.4).

Moreover, the VESs also provide useful guide for designing new bending-immune waveguides. The most intuitive design is perhaps to interface a valley-topological PhC (Fig.4a) with a trivial counterpart (Fig.4d). Both PhCs have **K/K'** valleys in the bulk bands, so that the VESs with momentum hotspots at the **K** or **K'** points could be supported. Figure 4c confirms this scenario and demonstrates a Z-shaped bending immune transmission. Figure 4e demonstrates another design by utilizing a topological invariant named 2D Zak phase, which classifies 2D crystals according to the localization of Wannier center. In this design, the upper and lower PhCs are both valley-topological trivial (with zero valley-Chern number). Nevertheless, they are judiciously cut along the zig-zag interface to generate contrasting distinct 2D Zak phases, [0,0] and [1/2, 1/2], so that the in-gap edge states emerge (see supplementary Sec.6). The modal profile suggests that the exemplified edge state belongs to the VESs and a Z-shaped transmission simulation confirms its bending immunity (Fig.4e). We want to stress that the 2D Zak phase itself does not guarantee the bending immunity (see supplementary Sec.6), instead, this property only occurs when the generated edge states are VESs.

VESs could be extended beyond $C_3$ symmetric systems. For PhCs made of rhombic lattices, of which the reciprocal lattice can be regarded as deformed from a regular hexagon, the six vertices of the first Brillouin zone can be similarly divided into two classes, denoted by $\mathbf{O, M, \overline{N}}$ or $\mathbf{\overline{O}, \overline{M}, N}$ points (Fig.4f). As long as an incident edge state has momentum hotspots at $\mathbf{O, M, \overline{N}}$ ($\mathbf{\overline{O}, \overline{M}, N}$) points, the bending immunity can be preserved as a result of the same mechanism in the hexagonal lattice. Figure 4f demonstrates a high transmittance bending in a case where the sharp angle of the rhombic lattice is 30° (the bending angle is thus 150°). Note that, in this case, due to that the bulk valleys slightly deviates from the vertex points (see supplementary Sec.7), the momentum hotspots are not exactly at the $\mathbf{O, M, \overline{N}}$ ($\mathbf{\overline{O}, \overline{M}, N}$) points. For this reason, such edge states are called quasi-VESs.

**Conclusion**

In this letter, we clarify that unidirectional transmission against 120° sharp bending and random disorder is not the privilege of valley-topological photonics. A unique modal character of the edge states, with momentum hotspots around high-symmetric **K** (**K'**) points, is responsible for bending immunity. We refer this specific type of bending-immune edge modes as valley edge states (VESs), which offers a coherent interpretation for bending immunity in a range of topological and topological-trivial edge states implemented in $C_3$ crystals, and also provides intuitive guide for designing new bend-immune waveguides, e.g., employing interfaced PhCs of rhombic lattices.

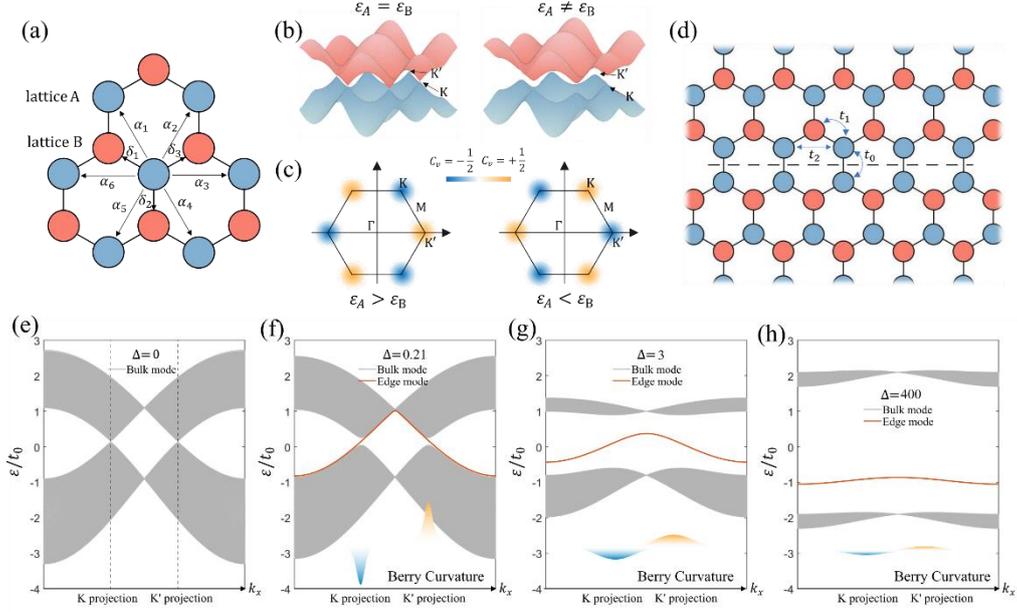

**Figure 1. Generation and evolution of valley-topological states in tight-binding model.** (a) A hexagonal lattice constructed by sublattice A (blue) and B (red) with energy $\varepsilon_A$ and $\varepsilon_B$. $\boldsymbol{\delta_i}$ ($i = 1,2,3$) and $\boldsymbol{\alpha_i}$ ($i = 1, 2, \ldots 6$) represent the nearest and next-nearest hopping vectors. (b) Band diagrams with (left) and without (right) **P**-symmetry. (c) A tiny **P**-symmetry breaking generates opposite Berry curvatures localized at **K** and **K'** points in the valence band, corresponding to $C_v = \pm 1/2$ [similarly to (f)]. (d) Two hexagonal crystals, with A, B sublattices swapped, are interfaced along the zig-zag direction to generate edge states with hopping parameters: $t_1$, nearest A-B hooping energy; $t_2$, next nearest A-A (B-B) hooping energy; $t_0$, A-A coupling energy across the interface. (e)-(h) Band diagrams of bulk (gray) and edge states (red) at different detuning ratios $\Delta \equiv (\varepsilon_B - \varepsilon_A)/t_1 = 0, 0.21, 3, 400$. The hopping parameters are set as $t_1 = t_0, t_2 = 0.05t_0$ (e); $t_1 = 0.95t_0, t_2 = 0.05t_0$ (f); $t_1 = 0.5t_0, t_2 = 0.05t_0$ (g); $t_1 = 0.01t_0, t_2 = 0.05t_0$ (h). The insets plot Berry curvature. With a small detuning ratio e.g., $\Delta = 0.21$, the Berry curvatures are localized at the **K** and **K'** points, and gapless valley-topology edge states emerge. As the detuning ratio increases, the valley-topology weakens and the topological edge states gradually turn to trivial edge ones.

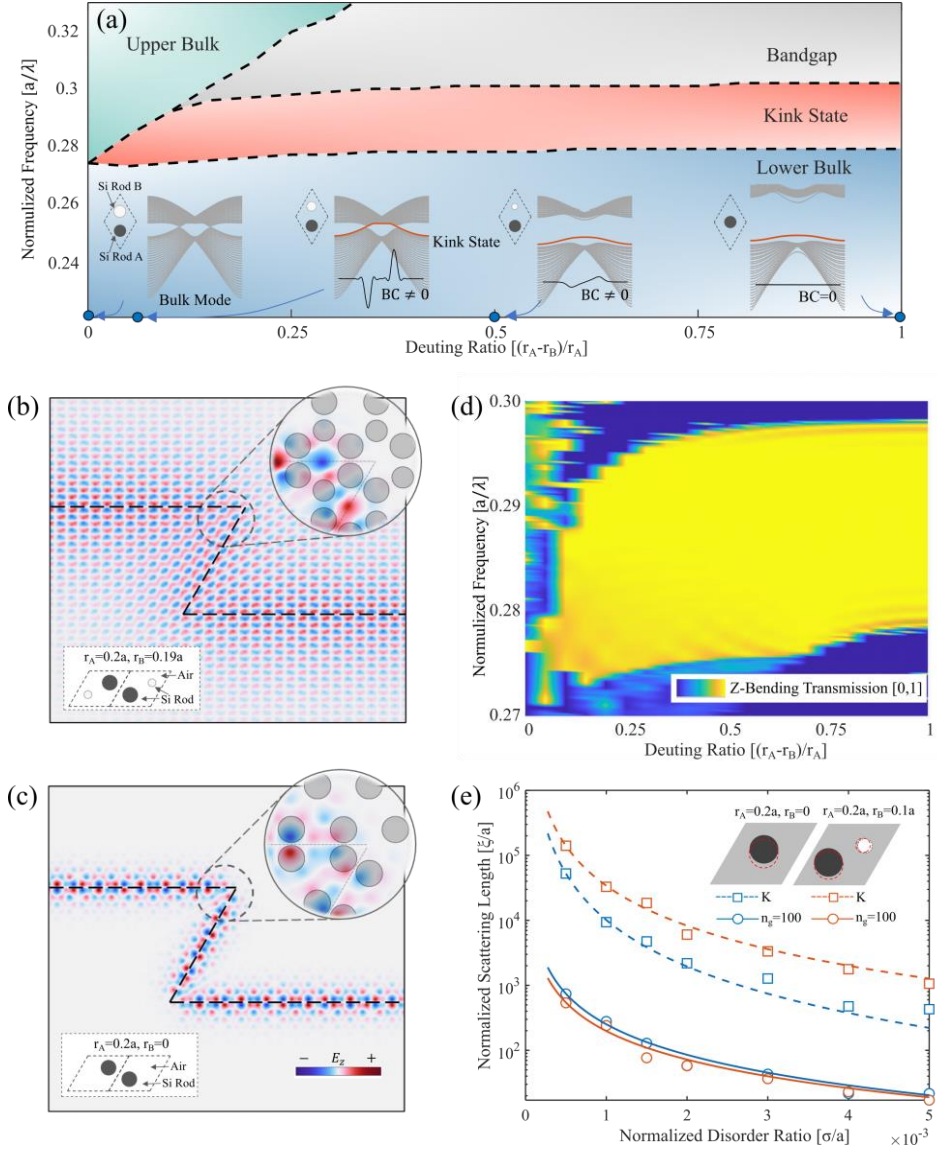

**Figure 2. Evolution of photonic valley-topological edge states and their universal propagation robustness during a continuous P-symmetry breaking procedure**. A hexagonal photonic lattice (lattice constant a) is composed of Si rods A (radius $r_A$) and B (radius $r_b$) in air background [see the insets in (a)]. The edge states ($E_z$-polarized) are generated on the zig-zag interface between two PhCs with A, B switched. (a). Bandwidth of lower, upper bulk bands and edge states—versus detuning ratio $(r_A - r_B)/r_A$ that reflects P-symmetry breaking. $r_A$ is fixed to $0.2a$. The insets sketch the lattice configurations and plot the projected band diagrams for detuning ratios of 0, 0.05, 0.5 and 1. (b) and (c) Distributions of $E_z$ for edge states in Z-bending waveguides composed of topological (detuning ratio 0.05) and trivial PhCs (detuning ratio 1). The wavenumber is set to $2\pi/3a$. (d) Z-bending Transmission spectra of edge states within PhCs of different detuning ratios. (e) Normalized scattering length via random disorder ratio in straight PhC waveguides with detuning ratio 0.5 (red) and 1 (blue). Round and square marks correspond to projected **K** point and group index $n_g = 100$, respectively, and dashed and solid lines are their fitted curves.

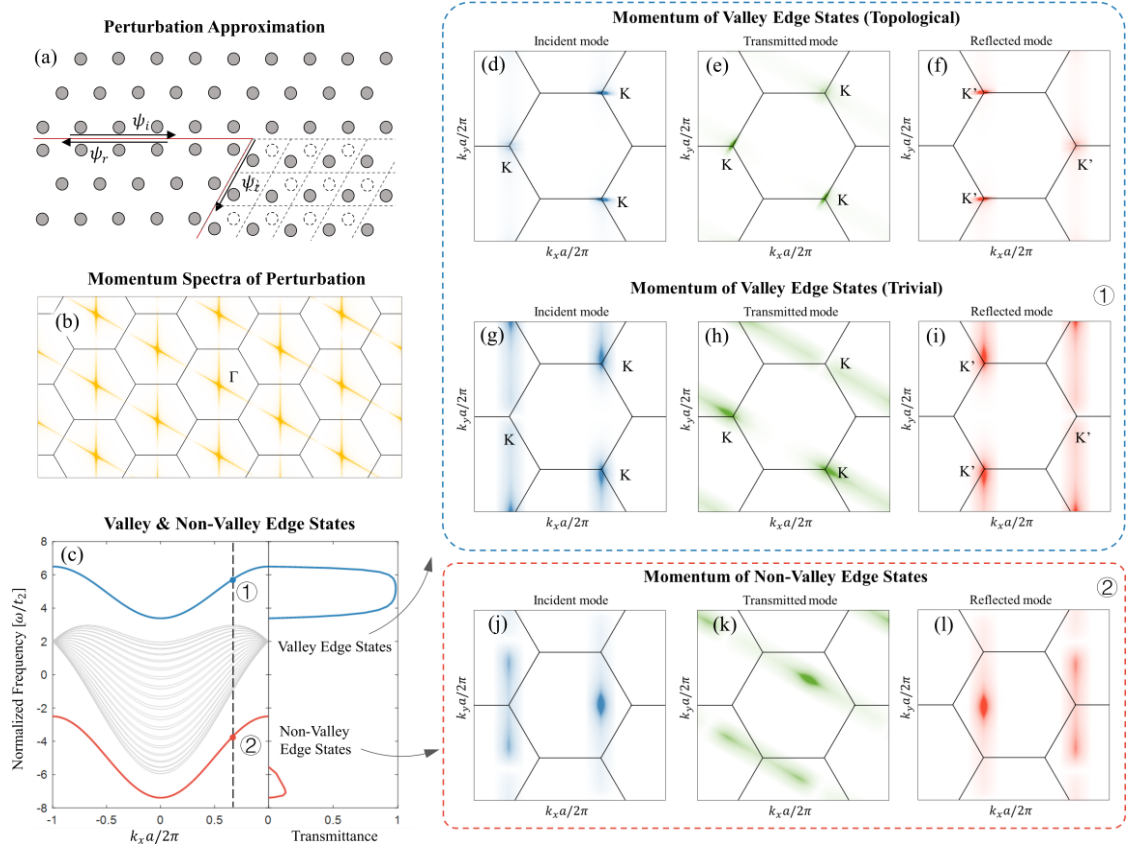

**Figure 3. Momentum matching analysis for bending immunity.** (a) Sketch of wave scattering in a bending waveguide. The scatters are parameterized by the permittivity redistribution from straight to bending waveguide (position shift of Si rods from dashed circles to solid ones). The incident, reflected, transmitted edge states are denoted by $\psi_i$, $\psi_r$ and $\psi_t$. (b) Momentum profile of approximated point scatters, $|\Delta\varepsilon(\boldsymbol{k})|$. The upper-limit of the $|\Delta\varepsilon(\boldsymbol{k})|$ is adjusted to make the bright lines more recognizable. (c) Projected band diagrams of zig-zag interfaced valley-topology trivial crystals (sublattice B eliminated, $\varepsilon_A = 0$, and $t_2 = t_0/3$) described by the tight-binding model and bending transmittance of upper (valley) and lower (non-valley) branches of edge states. (d)-(i), (j)-(l) Momentum profiles of incident (blue), transmitted (green) and reflected (red) modes corresponding to the projected **K** points for valley and non-valley edge states. In (g)-(i) and (j)-(l), the incident edge states correspond to the modes ① and ② labelled in (c), respectively. (d)-(f) considers a valley-topological case with a tiny AB detuning, in which the tight-binding parameters are $\Delta \equiv (\varepsilon_B - \varepsilon_A)/t_1 = 0.2$, $t_1 = t_0$, $t_2 = 0$.

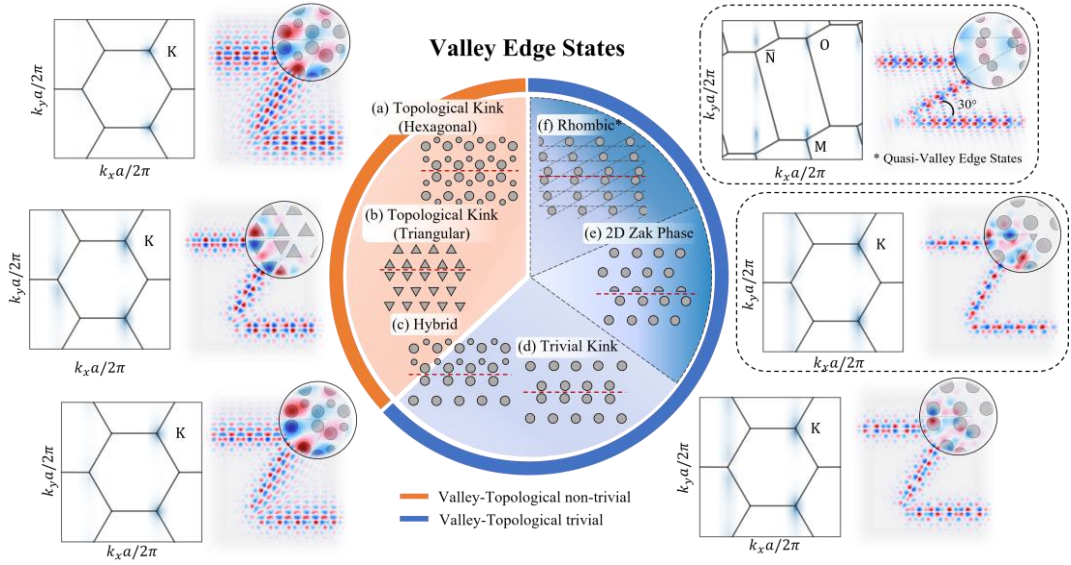

**Figure 4. Valley edge states (VESs), a class of edge states that supports bending immunity.** Bending-immune VESs are demonstrated with a set of examples implemented in $C_3$ symmetric crystals, including valley-topological (VT) cases (a) and (b), hybrid VT trivial-non-trivial cases (c), VT trivial cases (d), and VT trivial but topological non-trivial 2D-Zak-phase cases (e). The implementation of VESs can also be generalized to rhombic lattices, see 30° rhombic lattice in (f). Momentum profiles of edge states and their perfect transmission through a Z-shaped waveguide are shown for all exemplified cases (See supplementary Sec.8 for simulation details).

# SUPPLEMENTARY INFORMATION
## S1. Numerical Calculations for the Universal Propagation Properties

In the Fig. 2 of the main text, we demonstrate the propagation properties of the valley topological (trivial) edge states. The two-dimensional calculations are performed with the finite element method (FEM) using COMSOL Multiphysics. The calculation details are presented below.

### S1.1 Z-shaped Bending Calculation

The construction of Z-shaped bending waveguides is illustrated in Fig. S1.1(a), where the lattice constant is set to be $a = 500$ nm, the radius of sub-lattice A ($r_A$) and B ($r_B$) is set to be $r_A = 200$nm, $r_B = (1-D)r_A$ respectively. The intermediate section between two bending is set to be 16-lattice long. Notice that the interface between the PhCs, so as to say, the termination of either PhC, is aligned with the zig-zag direction of the lattice (marked with solid lines), hence no lattice is cut in the system.

The propagating modes ($E_z$-polarized) are stimulated by an in-plane magnetic dipole, the orientation of which is aligned with the zig-zag interface. Both forward- (right side of the dipole) and backward- (left side of the dipole) propagating modes can be stimulated. The power flow of the transmitted mode (marked as $P_{bending}$) is obtained by integrating the Poynting vector cross the monitor surface. The dimension of the monitor surface is set larger than the lateral confinement length of the modes.

To avoid the reflection of either forward- or backward- propagating modes, the region of interest is terminated by two gradual absorption regions which work as nearly-perfect reflection free terminations. In the gradual absorption region, the refractive indices of Si rods are set to be $3.4 + i \cdot \alpha |d|^2$, where $|d|$ represents the *x*-direction distance away from the lossless region and $\alpha$ is a coefficient. In the calculation corresponding to Fig. 2 of the main text, the absorption region is set to be 30 lattice long and $\alpha$ is set to be $3 \times 10^8 \, [m^{-2}]$.

To get the transmittance ratio, an accompanied straight waveguide is formed, the parameters of which are the same as the Z-shaped waveguide and the monitored power flow ($P_{straight}$) is used as a reference (Fig. S1.1(b)). The transmittance demonstrated in Fig. 2(d) is calculated by $P_{bending}/P_{straight}$. Notice that such calculation will encounter problems when group index $n_g \to \infty$. Because in this case, both the numerator ($P_{bending}$) and the denominator ($P_{straight}$) approach 0. More specifically, such cases appear around the cutoff frequencies of the edge states where the group index is larger than 150 as observed in our calculation. However, since the affected spectral region is negligible, plus we care more about the passband of high transmittance, this calculation error is tolerable. In Fig. 2(d), we simply set a 100% transmittance cut-off to block the abnormal data.

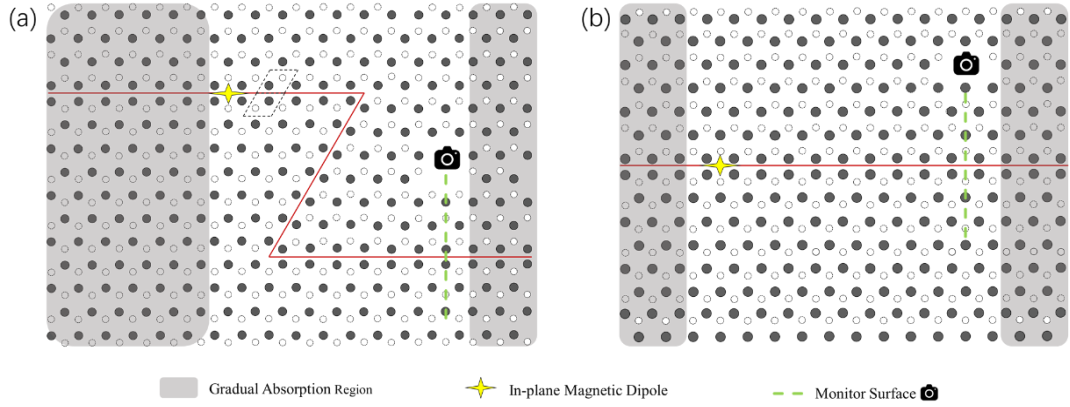

**Fig S1.1. Simulation settings for Z-shaped waveguide (a) and straight waveguide (b).** In both cases, the propagating modes are stimulated from in-plane magnetic dipoles (yellow stars). The power flows are detected at the monitor surface (dashed green lines). The lossless region of interest is terminated with two gradual absorption regions (shadowed) as equivalent reflection-free terminations.

### S1.2 Random Disorder Calculation

The transport simulation with random disorder is performed in a straight waveguide similar to Fig. S1.1(b), with the only difference being adding a 50-lattice ($50a$) long random disorder region between the source (dipole) and the monitor, as shown in Fig. S1.2. Drawing on the previous literature[1], the random disorder is introduced by setting the position of the Si rods according to a 2D normal distribution $N(\boldsymbol{\mu}, \sigma^2)$. The 2D probability density function for the position of the Si rods is $f(\boldsymbol{r}) = 1/(2\pi\sigma^2) e^{-1/2 \cdot |\boldsymbol{r}-\boldsymbol{\mu}|^2/\sigma^2}$, where $\boldsymbol{\mu} = (x_0, y_0)$ is the mean position of the randomized Si rods and $\sigma$ is the standard deviation.

We measure the performance of the waveguides against random disorder with a parameter named back scattering length ξ. Similar to the concept of scattering mean free path in condensed matter physics, ξ measures the propagation length after which the light will be noticeably scattered. The ξ here is standardized to be the value which makes the power flow attenuate to 1/e of its original value, as expressed in Eq. S1,

$$T(x) = \frac{I_\sigma}{I_0} = e^{-\frac{x}{\xi}}. \qquad \text{Eq. S1}$$

Since the random disorder region is $50a$ long in the $x$-direction, i.e., $x = 50a$ in Eq. S1, the back scattering length is calculated as $-50a/\ln(I_\sigma/I_0)$, where $I_\sigma$ and $I_0$ are the monitored power flow with and without random disorder. Naturally, the calculated backscattering length corresponding to each generated random sequence is different. To get a mean backscattering length, we average each data point in Fig. 2(e) over 60-time calculations.

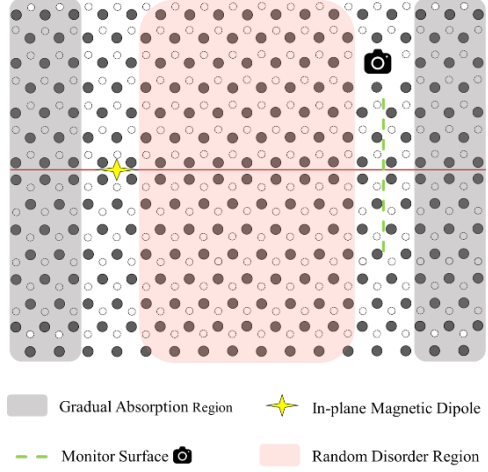

**Fig. S1.2. Simulation settings for straight waveguides with random disorders.** The propagating modes are stimulated from an in-plane magnetic dipole marked with yellow star. The random disorder is introduced by perturbing the positions of Si rods according to a normal distribution in the random disorder region (pink). The power flows are detected at the monitor surface (dashed green line). The region of interest is terminated with two gradual absorption regions (shadowed).

## S2. Numerical calculations of the reflection and transmission measures

In the main text, we describe a topological-irrelevant perspective, which regards the bending problem as a scattering problem and describe the scattering into reflection or transmission modes with a 1st order perturbation approximation. By performing Fourier transformations to the edge states and the scatters, and further analyzing their momentum profiles, we find that the specific momentum pattern with localized hotspots on **K** or **K'** points results in a significant difference in the overlap integrals between incident & transmitted modes and incident & reflected modes. We named edge states with this property as valley edge states and put forward such concept to interpret bending immunity. In this section, we will provide complementary discussions in the real space and associated numerical results.

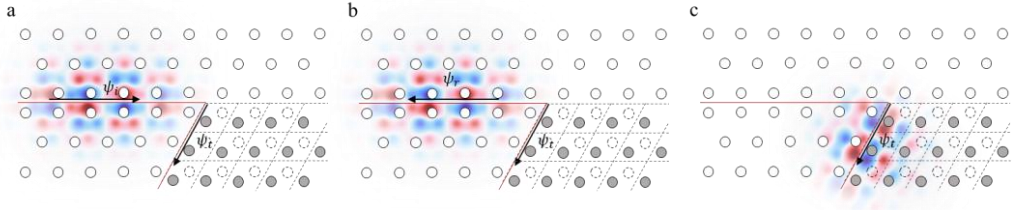

**Fig. S2.1. Real-space modal profiles of incident, reflected and transmitted modes.**

The expression of the electric fields of the incident edge state is

$$\varphi_{\vec{k}} = u_1^{\vec{k}}(r)e^{ikx} \cdot e^{-\kappa_1 y}{}_{(y>0)} + u_2^{\vec{k}}(r)e^{ikx} \cdot e^{\kappa_2 y}{}_{(y<0)} , \qquad \text{Eq. S2.1a}$$

where $\kappa_1$ and $\kappa_2$ are the constants describing the lateral confinement. The field expression consists of periodic term $u^{\vec{k}}(r)$ with the same periodicity as the lattice (i.e., $u^{\vec{k}}(r) =$ $u^{k_\rightarrow}(r - ma_1 - na_2))$, propagating term $e^{ikx}$ and lateral confinement terms $e^{-\kappa_1 y}$ and $e^{\kappa_2 y}$. The difference between this expression and the Bloch expansion lies in the extra confinement term (see Fig. S2.1), which has also been used in Ref. [2]. Similarly, we can write the electric fields of the reflected and transmitted edge states as

$$\varphi_{k_\leftarrow} = u_1^{k_\leftarrow}(r)e^{-ikx} \cdot e^{-\kappa_1 y}\Big|_{(y>0)} + u_2^{k_\leftarrow}(r)e^{ikx} \cdot e^{\kappa_2 y}\Big|_{(y<0)},  \quad \text{Eq. S2.1b}$$

$$\varphi_{k_\downarrow} = u_1^{k_\downarrow}(r)e^{ikx'} \cdot e^{-\kappa_1 y'}\Big|_{(y'>0)} + u_2^{k_\downarrow}(r)e^{ikx'} \cdot e^{\kappa_2 y'}\Big|_{(y'<0)},  \quad \text{Eq. S2.1c}$$

where $x' \equiv -x\cos(\theta) - y\sin(\theta)$ and $y' \equiv x\sin(\theta) - y\cos(\theta)$ with $\theta = \pi/3$.

The permittivity redistribution is given by

$$\Delta\varepsilon(r) = \int dr' f(r') \cdot g(r') \cdot h(r - r'). \quad \text{Eq. S2.2}$$

Here $f(r) = \sum_{m,n} \delta(r - ma_1 - na_2)$ describes the lattice periodicity, and $a_1 = (a, 0)$ and $a_2 = (a/2, \sqrt{3}a/2)$ are lattice vectors. $g(r) = 1$ in the perturbation region (red region in Fig. S2.2) and $g(r) = 0$ otherwise (blue region in Fig. S2.2). $h(r)$ describes the profile of the permittivity redistribution in a single lattice.

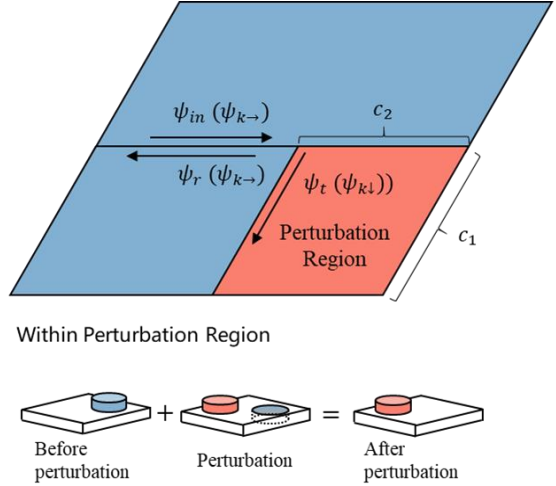

**Fig. S2.2. The illustration of perturbation region and edge states.** The rhombic perturbation region is dyed with red. Within each lattice of the perturbation region, the permittivity is redistributed, as illustrated in the lower panel.

The measures of transmission and reflection are

$$A_t = \int_{-\infty}^{+\infty}\int_{-\infty}^{+\infty} dx \cdot dy \cdot \varphi_{k_\rightarrow}^*(r)\varphi_{k_\downarrow}(r) \cdot \Delta\varepsilon(r), \quad \text{Eq. S2.3(a)}$$

$$A_r = \int_{-\infty}^{+\infty}\int_{-\infty}^{+\infty} dx \cdot dy \cdot \varphi_{k_\rightarrow}^*(r)\varphi_{k_\leftarrow}(r) \cdot \Delta\varepsilon(r). \quad \text{Eq. S2.3(b)}$$

By defining the integration range in the perturbation region (i.e., $y < 0$ and $x\sin\theta > y\cos\theta$) where $\Delta\varepsilon(\mathbf{r})$ is nonzero, the integrals in Eqs. S2.3 are derived as

$$A_t = \sum_{m,n} \int_{-\infty}^{0} dy \int_{y/\tan\theta}^{\infty} dx \cdot u_2^{k_\rightarrow *}(\mathbf{r}) u_1^{k_\downarrow}(\mathbf{r}) \cdot e^{i(-k\cdot x - k\cos\theta\cdot x - k\sin\theta\cdot y)} \cdot e^{\kappa_2\cdot y - \kappa_1\cdot(\sin\theta x - \cos\theta y)} \cdot h(\mathbf{r} - m\mathbf{a}_1 - n\mathbf{a}_2)$$

$$= F_{k_\rightarrow, k_\downarrow} \sum_{m>0, n<0} e^{i(-k\cdot x_{mn} - k\cos\theta\cdot x_{mn} - k\sin\theta\cdot y_n)} \cdot e^{\kappa_2\cdot y_n - \kappa_1\cdot(\sin\theta x_{mn} - \cos\theta y_n)}, \qquad \text{Eq. S2.4(a)}$$

$$A_r = \sum_{m,n} \int_{-\infty}^{0} dy \int_{y/\tan\theta}^{+\infty} dx \cdot u_2^{k_\rightarrow *}(\mathbf{r}) u_2^{k_\leftarrow}(\mathbf{r}) e^{-i2k\cdot x} \cdot e^{2\kappa_2\cdot y} \cdot \delta(x - x_{mn})\delta(y - y_n)$$

$$= F_{k_\rightarrow, k_\leftarrow} \sum_{m>0, n<0} e^{-i\cdot 2k\cdot x_{mn}} \cdot e^{2\kappa_2\cdot y_n}, \qquad \text{Eq. S2.4(b)}$$

with

$$F_{k_\rightarrow, k_\downarrow} \equiv \int u_2^{k_\rightarrow *}(\mathbf{r}) u_1^{k_\downarrow}(\mathbf{r}) h(\mathbf{r}) e^{i(-k\cdot x - k\cos\theta\cdot x - k\sin\theta\cdot y)} \cdot e^{\kappa_2\cdot y - \kappa_1\cdot(\sin\theta x - \cos\theta y)} d\mathbf{r}, \qquad \text{Eq. S2.4(c)}$$

$$F_{k_\rightarrow, k_\leftarrow} \equiv \int u_2^{k_\rightarrow *}(\mathbf{r}) u_1^{k_\leftarrow}(\mathbf{r}) h(\mathbf{r}) e^{-i2k\cdot x} \cdot e^{2\kappa_2\cdot y} d\mathbf{r}, \qquad \text{Eq. S2.4(d)}$$

$x_{mn} \equiv \left(m + \frac{n}{2}\right)a$ and $y_n \equiv \frac{\sqrt{3}}{2} na$. Note that, the overlap integrals $F_{k_\rightarrow, k_\downarrow}$ and $F_{k_\rightarrow, k_\leftarrow}$ are performed in the single lattice where $h(\mathbf{r})$ is defined. Moreover, the evaluations of the reflection and transmission measure only concern the edge states in the perturbation region. Therefore, unlike the main text wherein the momentum profiles of the edge states in the whole space are discussed, in the supplementary document hereafter, only the edge states in the relevant perturbation region (i.e., $\varphi_{k_\rightarrow, y<0}$, $\varphi_{k_\leftarrow, y<0}$, $\varphi_{k_\downarrow, \sin\theta x > \cos\theta y}$) are considered.

Notice that in Eqs. S2.4, the field overlap terms $F_{k_\rightarrow, k_\downarrow}$ and $F_{k_\rightarrow, k_\leftarrow}$ are extracted from the series summation since they are independent of lattice indexes $(m,n)$. Accordingly, $A_t$ and $A_r$ are contributed from two decoupled terms, the field overlap term and the summation of the propagation terms. The first term is tightly bounded with specific periodic modal and perturbation profile. According to a myriad of numerical observations, the bending-immune property consistently appears in a variety of edge states with distinct periodic modal and perturbation features. This indicates the propagation term plays the key role in general.

Regarding the propagation term, we first notice that it possesses an interesting character when $k = \pm 4\pi/3a$ corresponding to the **K'** and **K** points (the lateral confinement parameters $\kappa_1$ and $\kappa_2$ are assumed to be positive real-valued). Specifically, for the transmission measure $A_t$, the lattice sampled terms $e^{i(-k\cdot x_{mn} - k\cos\theta\cdot x_{mn} - k\sin\theta\cdot y_n)}$ are equal to 1 independent of $m$ and $n$ (see Fig. S2.3), thus leading to a constructive summation. This is simply because that difference of the propagation wavevectors between the incident and transmission edge states (that is $\mathbf{k}_\rightarrow = k\hat{x}$ and $\mathbf{k}_\downarrow = -k/2\hat{x} - \sqrt{3}k/2\hat{y}$, respectively, and $k = \pm 4\pi/3a$) equals the reciprocal lattice vector. Contrastingly, for the reflection measure, the involved propagation terms $e^{-i\cdot 2k\cdot x_{mn}}$ have $\pm 4\pi/3a$ and $\pm 2\pi/3a$ phase differences between neighboring lattices $m$ & $m+1$ and $n$ & $n+1$, which results in a destructive summation.

This perspective provides an intuitive understanding why the phase summation for the transmission measure is preponderant compared with that of the reflection measure. This real-space analysis echoes with the k-space analysis demonstrated in the main text, and they describe the same bending mechanism from two complementary perspectives.

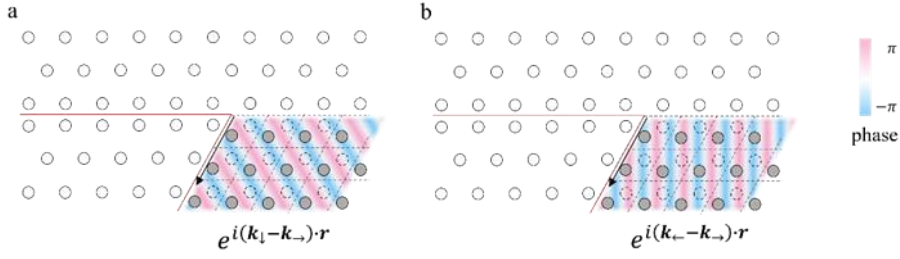

**Fig. S2.3. The phase distribution of propagation terms in the transmission (a) and reflection (b) measures.** The color represents the phase from $-\pi$ to $\pi$. The wavevectors of the incident, reflection and transmission edge states are set to be $\boldsymbol{k}_\rightarrow = k\hat{x}$, $\boldsymbol{k}_\leftarrow = -k\hat{x}$ and $\boldsymbol{k}_\downarrow = -k/2\hat{x} - \sqrt{3}k/2\hat{y}$ with $k = 4\pi/3a$.

The simplified transmission/reflection ratio, only including the summation of the propagation terms, are calculated. When the wavevector is set corresponding to **K'** point (gray dashed line in Fig. S2.4), according to the modified Bloch expansion, the momentum profile of the incident mode will be localized at **K'** points. As demonstrated in Fig. S2.4, a low reflection (high transmission) band appears around this wavevector. The different-color curves correspond to different values of lateral confinements with $D$ marking the decay length (lateral distance where the field amplitude decays to 1/e of its maximum value). It can be observed that the passband is robustly present when D varies from $20a$ to $3a$, which confirms that such bending-immune property is insensitive to the lateral confinement.

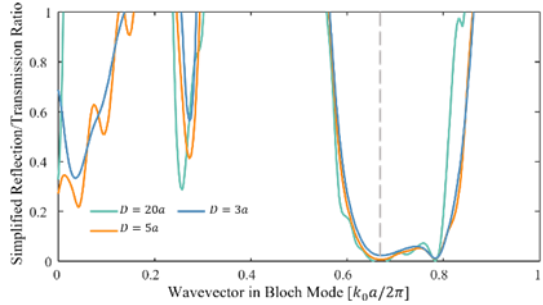

**Fig. S2.4. The numerical calculations of reflection/transmission ratio via wavevector for propagating modes with different decay length $D$.** Low reflection (high transmission) region is observed around the wavevector corresponding to **K'** points (gray dashed line). In the simulation, the lateral confinement parameters are set to be $\kappa_1 = \kappa_2 = 1/D$.

## S3. Analytic Analysis of Momentum Profiles of Edge States with Tight-Binding Model

To provide more insights into momentum profiles of edge states that lead us to the definition of valley edge states (VESs), we perform analytic analysis within the tight-binding model. The tight-binding model considers two zig-zag-interfaced hexagonal lattices with A and B sublattice (see Fig. S3.1 and the related descriptions in the main text for more details). Further, to simplify our analysis, we study two representative cases: (i) topological-trivial case with sublattice B eliminated, see Fig. S3.1(a); (ii) valley topological case with next-nearest (AA and BB) hoping omitted, see Fig. S3.1(b). The analysis below is used to supplement the numerical results of Figs. 3(d)-(i) in the main text.

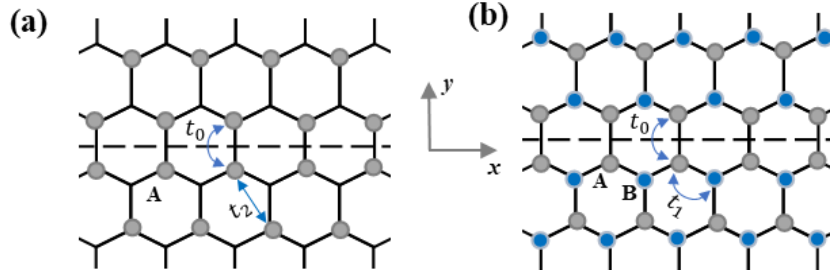

**Fig. S3.1. Schematics of topological-trivial (A) and valley topological (B) systems in a tight-binding model.** The hopping parameters $t_1$, $t_2$ and $t_0$ are labelled.

### S3.1 Topological-trivial case

**Dispersion Relations.** — In the topological-trivial case, the sublattice B is removed [Fig. S3.1(a)]. The resonant energy of the sublattice A is set to be zero. The complex-valued amplitudes of the atoms A are denoted by $\phi(r_{+i})$ with $y > 0$ (the upper half space) and $\phi(r_{-i})$ with $y < 0$ (the lower half space), respectively, with

$$\phi(r_{+i}) = A_+ \exp(i\beta x_{+i}) \exp(-\alpha y_{+i}), \qquad Eq.\,S3.1(a)$$

$$\phi(r_{-i}) = A_- \exp(i\beta x_{-i}) \exp(\alpha y_{-i}), \qquad Eq.\,S3.1(b)$$

where $r_{\pm i} \equiv x_{\pm i}\hat{x} + y_{\pm i}\hat{y}$ ($i = 1,2,3, \ldots$) denote the atom coordinates. Using the hopping relations between the neighboring atoms, we have

$$\omega A_+ = -2t_2 A_+ \cos(\beta a) - 2t_2 A_+ \cos\left(\frac{\beta a}{2}\right)\exp\left(-\frac{\alpha\sqrt{3}a}{2}\right) - t_0 A_-, \qquad Eq.\,S3.2(a)$$

$$\omega A_- = -2t_2 A_- \cos(\beta a) - 2t_2 A_- \cos\left(\frac{\beta a}{2}\right)\exp\left(-\frac{\alpha\sqrt{3}a}{2}\right) - t_0 A_+, \qquad Eq.\,S3.2(b)$$

$$\omega A_+ = -2t_2 A_+ \cos(\beta a) - 2t_2 A_+ \cos\left(\frac{\beta a}{2}\right)\exp\left(-\frac{\alpha\sqrt{3}a}{2}\right)$$
$$-2t_2 A_+ \cos\left(\frac{\beta a}{2}\right)\exp\left(\frac{\alpha\sqrt{3}a}{2}\right), \qquad Eq.\,S3.2(c)$$

$$\omega A_- = -2t_2 A_- \cos(\beta a) - 2t_2 A_- \cos\left(\frac{\beta a}{2}\right)\exp\left(-\frac{\alpha\sqrt{3}a}{2}\right)$$
$$-2t_2 A_- \cos\left(\frac{\beta a}{2}\right)\exp\left(\frac{\alpha\sqrt{3}a}{2}\right), \qquad Eq.\,S3.2(d)$$

where $a$ is the lattice constant. Equations S3.2.(a) and S3.2.(b) express the hopping between the atoms across the zig-zag interface, while Eqs. S3.2.(c) and S3.2.(d) are for the bulky atoms.

Since the system has a parity symmetry with respect to the zig-zag interface, the edge states can be classified into the symmetrical and anti-symmetrical states, with $A_+ = A_-$ and $A_+ = -A_-$ differently. Imposing $A_+ = \pm A_-$ in Eqs. S3.2, the dispersion relations of the edge states are derived:

$$\omega = \left(-2t_2 - 2\frac{t_2^2}{t_0}\right)\cos(\beta a) - 2\frac{t_2^2}{t_0} - t_0 \text{ (symmetric)}, \quad \text{Eq. S3.3(a)}$$

$$\omega = \left(-2t_2 + 2\frac{t_2^2}{t_0}\right)\cos(\beta a) + 2\frac{t_2^2}{t_0} + t_0 \text{ (anti}-\text{symmetric)}. \quad \text{Eq. S3.3(b)}$$

The validity of the derived dispersion relations is numerically confirmed in Fig. S3.2. It is demonstrated that the analytically computed dispersion relations (circles) agree exactly with the numerical results (lines). Further, we detive that the decaying wavenumber $\alpha$ satisfies

$$t_0 = 2t_2 \cos\left(\frac{\beta a}{2}\right) \exp\left(\frac{\alpha\sqrt{3}a}{2}\right) \text{ (symmetric)}, \quad \text{Eq. S3.4(a)}$$

$$t_0 = -2t_2 \cos\left(\frac{\beta a}{2}\right) \exp\left(\frac{\alpha\sqrt{3}a}{2}\right) \text{ (anti}-\text{symmetric)}. \quad \text{Eq. S3.4(b)}$$

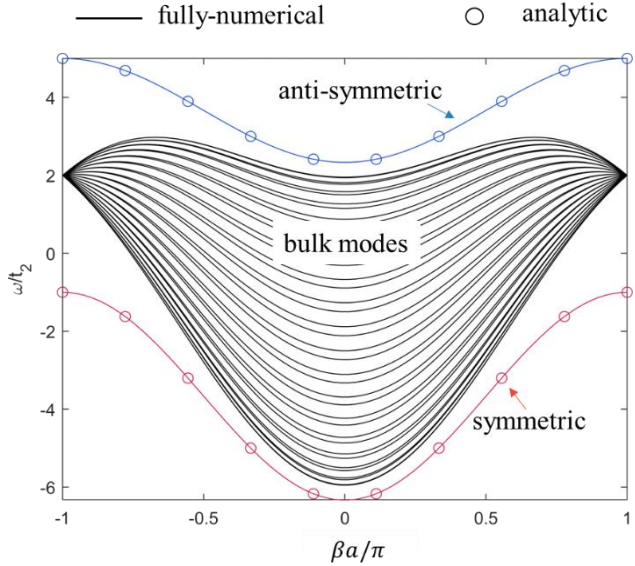

**Fig. S3.2. Numerical validation of derived dispersion relations [Eqs.S3.3] for topological-trivial edge states.** In the tight-binding model, we set $t_0 = 3t_2$.

Assuming that $t_0$ and $t_2$ have the same sign, the decaying wavenumber $\alpha$ is expressed as

$$\alpha = \alpha_{\beta1} \text{ (symmetric)}, \quad \text{Eq. S3.5(a)}$$

$$\alpha = \alpha_{\beta1} + i\alpha_{\beta2} \text{ (anti}-\text{symmetric)}, \quad \text{Eq. S3.5(b)}$$

with

$$\alpha_{\beta 1} \equiv \frac{2 \log\left(\frac{t_0}{2t_2 \cos\left(\frac{\beta a}{2}\right)}\right)}{\sqrt{3}a} \qquad \text{Eq. S3.6(a)}$$

and

$$\alpha_{\beta 2} \equiv \frac{2\pi}{\sqrt{3}a}. \qquad \text{Eq. S3.6(b)}$$

**Momentum Profiles.** — We then examine the modal profiles by performing the Fourier transformation to the atom amplitudes $\sum_i \phi(r_{\pm i})\delta(r - r_{\pm i})$, i.e.,

$$\sum_i \phi(r_{\pm i})\delta(r - r_{\pm i}) = \int dk \, \tilde{\phi}_{\pm}(k) \exp(ik \cdot r). \qquad \text{Eq. S3.7}$$

It is derived that

$$\tilde{\phi}_{\pm}(k) = \sum_n \sum_m \tilde{\phi}_{\pm}^0(k - G_{nm}) \exp(-iG_{nm} \cdot r_{\pm 1}), \qquad \text{Eq. S3.8(a)}$$

with

$$\tilde{\phi}_{\pm}^0(k) = \mp \frac{1}{2\pi S_{\text{lattice}}} A_{\pm} \frac{1}{\mp\alpha - ik_y} \delta(k_x - \beta) \text{ (symmetric)}, \qquad \text{Eq. S3.8(b)}$$

$$\tilde{\phi}_{\pm}^0(k) = \mp \frac{1}{2\pi S_{\text{lattice}}} A_{\pm} \frac{1}{\mp\alpha - ik_y} \delta(k_x - \beta) \text{ (anti-symmetric)}, \qquad \text{Eq. S3.8(c)}$$

$$G_{nm} = nb_1 + mb_2, \qquad \text{Eq. S3.8(d)}$$

where $S_{\text{lattice}}$ denotes the area of the hexagonal lattice, and $b_1 = 4\pi/(\sqrt{3}a)\hat{y}$ and $b_2 = 4\pi/(\sqrt{3}a)(\sqrt{3}/2\hat{x} + 1/2\hat{y})$ are the reciprocal lattice vectors.

Consider the edge states with wavenumber $\beta = \pm\frac{2\pi}{3a}$ corresponding to the projected **K** and **K'** points in the *x*-direction. When the edge states are anti-symmetric, from Eqs. S3.8, we see that they host hotspots centralized around $k = \pm\frac{2\pi}{3a}\hat{x} + \frac{2\pi}{\sqrt{3}a}\hat{y} + G_{nm}$, which are just **K** and **K'** points. On the other hand, for the symmetric ones, the momentum hotspots are shifted to $k = \pm\frac{2\pi}{3a}\hat{x} + G_{nm}$.

Apparently, in this case, the anti-symmetric edge states belong to the VESs, while the symmetric edge states do not.

When $t_0$ and $t_2$ have the opposite signs, by repeating the above routines, we find that the symmetric edge states instead become the VESs. Nevertheless, regardless of whether the anti-symmetric or symmetric edge states are the VESs which depends on the signs of the hopping parameters, it is always the edge states, which lie close to the bulk valleys in frequency, are the VESs (cf. Fig. S3.2). More discussions shall be provided in Sec. S4.

### S3.2 Valley topological case

**Dispersion Relations.** — In this case, we set the energies of the sublattice A and B to be $\epsilon_A = -\omega_0$ and $\epsilon_B = \omega_0$. The complex-valued amplitudes of the A and B atoms in the upper ($y > 0$) and lower

($y < 0$) half spaces separated by the zig-zag interface [Fig. S3.1(b)] are denoted by

$$\phi_A(\mathbf{r}_{A;+i}) = A_+ \exp(i\beta x_{A;+i}) \exp(-\alpha y_{A;+i}),$$  Eq. S3.9(a)

$$\phi_B(\mathbf{r}_{B;+i}) = B_+ \exp(i\beta x_{B;+i}) \exp(-\alpha y_{B;+i}),$$  Eq. S3.9(b)

$$\phi_A(\mathbf{r}_{A;-i}) = A_- \exp(i\beta x_{A;-i}) \exp(\alpha y_{A;-i}),$$  Eq. S3.9(c)

$$\phi_B(\mathbf{r}_{B;-i}) = B_- \exp(i\beta x_{B;-i}) \exp(\alpha y_{B;-i}).$$  Eq. S3.9(d)

We relate $A_\pm$ and $B_\pm$ with the hopping relations between the atoms and obtain that

$$\omega A_+ = -\omega_0 A_+ - 2t_1 B_+ \cos\left(\frac{\beta a}{2}\right) \exp\left(-\frac{\alpha a}{2\sqrt{3}}\right) - t_0 A_-,$$  Eq. S3.10(a)

$$\omega B_+ = \omega_0 B_+ - 2t_1 A_+ \cos\left(\frac{\beta a}{2}\right) \exp\left(\frac{\alpha a}{2\sqrt{3}}\right) - t_1 A_+ \exp\left(-\frac{\alpha a}{\sqrt{3}}\right),$$  Eq. S3.10(b)

$$\omega A_- = -\omega_0 A_- - 2t_1 B_- \cos\left(\frac{\beta a}{2}\right) \exp\left(-\frac{\alpha a}{2\sqrt{3}}\right) - t_0 A_+,$$  Eq. S3.10(c)

$$\omega B_- = \omega_0 B_- - 2t_1 A_- \cos\left(\frac{\beta a}{2}\right) \exp\left(\frac{\alpha a}{2\sqrt{3}}\right) - t_1 A_- \exp\left(-\frac{\alpha a}{\sqrt{3}}\right),$$  Eq. S3.10(d)

$$\omega A_+ = -\omega_0 A_+ - 2t_1 B_+ \cos\left(\frac{\beta a}{2}\right) \exp\left(-\frac{\alpha a}{2\sqrt{3}}\right) - t_1 B_+ \exp\left(\frac{\alpha a}{\sqrt{3}}\right),$$  Eq. S3.10(e)

$$\omega A_- = -\omega_0 A_- - 2t_1 B_- \cos\left(\frac{\beta a}{2}\right) \exp\left(-\frac{\alpha a}{2\sqrt{3}}\right) - t_1 B_- \exp\left(\frac{\alpha a}{\sqrt{3}}\right).$$  Eq. S3.10(f)

Solving Eqs. S3.10, the dispersion relations of the symmetric and anti-symmetric edge states are derived:

$$(\omega + \omega_0 + t_0)\left(\omega - \omega_0 + \frac{t_1^2}{t_0}\right) = 2t_1^2 \cos(\beta a) + 2t_1^2 \text{ (symmetric)},$$  Eq. S3.11(a)

$$(\omega + \omega_0 - t_0)\left(\omega - \omega_0 - \frac{t_1^2}{t_0}\right) = 2t_1^2 \cos(\beta a) + 2t_1^2 \text{ (anti} - \text{symmetric)}.$$  Eq. S3.11(b)

The validity of Eqs. S3.11 is numerically confirmed in Fig. S3.3 by setting $t_1 = t_0$ and $\omega_0 = 0.5t_1$.

The decaying wavenumber $\alpha$ can be computed with the following equations:

$$e^{-\frac{\sqrt{3}\alpha a}{2}} = \frac{-t_1 t_0 - t_1(\omega + \omega_0)}{2t_1 t_0 \cos\left(\frac{\beta a}{2}\right)} \text{ (symmetric)},$$  Eq. S3.12(a)

$$e^{-\frac{\sqrt{3}\alpha a}{2}} = \frac{-t_1 t_0 + t_1(\omega + \omega_0)}{2t_1 t_0 \cos\left(\frac{\beta a}{2}\right)} \text{ (anti} - \text{symmetric)}.$$  Eq. S3.12(b)

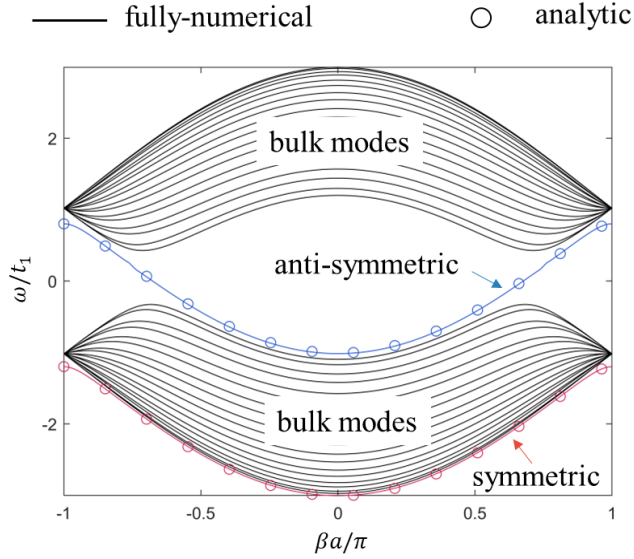

**Fig. S3.3. Numerical validation of derived dispersion relations [Eqs. S3.11] for valley topological edge states.** In the tight-binding model, we set $t_0 = t_1$ and $\omega_0 = 0.5t_1$.

**Momentum Profiles.** — As in the topological-trivial case, we can perform the Fourier transformation to the atom amplitudes $\sum_i \phi_A(r_{A;\pm i})\delta(r - r_{A;\pm i})$ or $\sum_i \phi_B(r_{A;\pm i})\delta(r - r_{A;\pm i})$, and examine the momentum profiles of the edge states. The Fourier transformation here takes the same form of Eqs. S3.8. Moreover, when the terms on the right-handed side of Eqs. S3.12 are negative (and note that $\beta \in (-\pi/a\ \pi/a)$), the decaying wavenumber $\alpha$ has a non-zero imaginary part with $\text{Im}(\alpha) = \frac{2\pi}{\sqrt{3}a}$, which is the same as in the topological-trivial case discussed above. In this case, it is straightforward to check with Eqs. S3.8 that the edge states, with wavenumber corresponding to the projected **K/K'** points, have momentum hotspots around **K/K'** points, and are thus VESs. However, it is tedious to derive under which explicit conditions that the imaginary part of the decaying wavenumber $\alpha$ is non-zero. Nevertheless, we numerically observe that it is always the edge states that lie between two bands (e.g., anti-symmetric edge states in Fig. S3.3), an argument for the validity of which is provided in Sec. S4.

In Fig. 3(c) in the main text, we have shown that, in the topological-trivial case, the VESs exhibit a high transmittance crossing a 120º bending, while the non-VESs do not. Below, we intend to valid that same conclusion also holds in the topological case. As is shown in Fig. S3.3, the anti-symmetric edge states are the VESs, while the symmetric edge states are not. However, in this case, the symmetric edge states are spectrally overlap with the bulk modes, so that they could be easily scattered into the latter, thereby lowering their bending transmittance. To avoid this undesirable overlap, we change the hooping parameters in Fig. S3.3 to $t_0 = 3t_1$. The projected-band diagram is plotted in the left panel of Fig. S3.4 (a), which shows three branches of the edge states that all have a certain frequency range that is completely separated from the bulk modes. The lower (red), middle (blue), and upper (green) branches correspond to symmetric, anti-symmetric and anti-symmetric edge states, respectively [see their representative real-space modal profiles in the rightmost panel of Fig. S3.4]. Moreover, the momentum profiles for the edge states with wavenumber $\beta = \frac{2\pi}{3a}$ corresponding to the projected **K** point are plotted in the left panel of Fig. S3.4 (b)-(d) verify that the edge states in the middle branch belong to the VESs. The simulated transmittance in the right panel of Fig. S3.4 (a) show that the VESs (blue) have a high transmittance close to 1, while the transmittance for the non-VESs (red and green) is strongly reduced. Additionally, note that, in the transmittance simulations, we only consider the frequency range where the edge states are spectrally separated with the bulk modes.

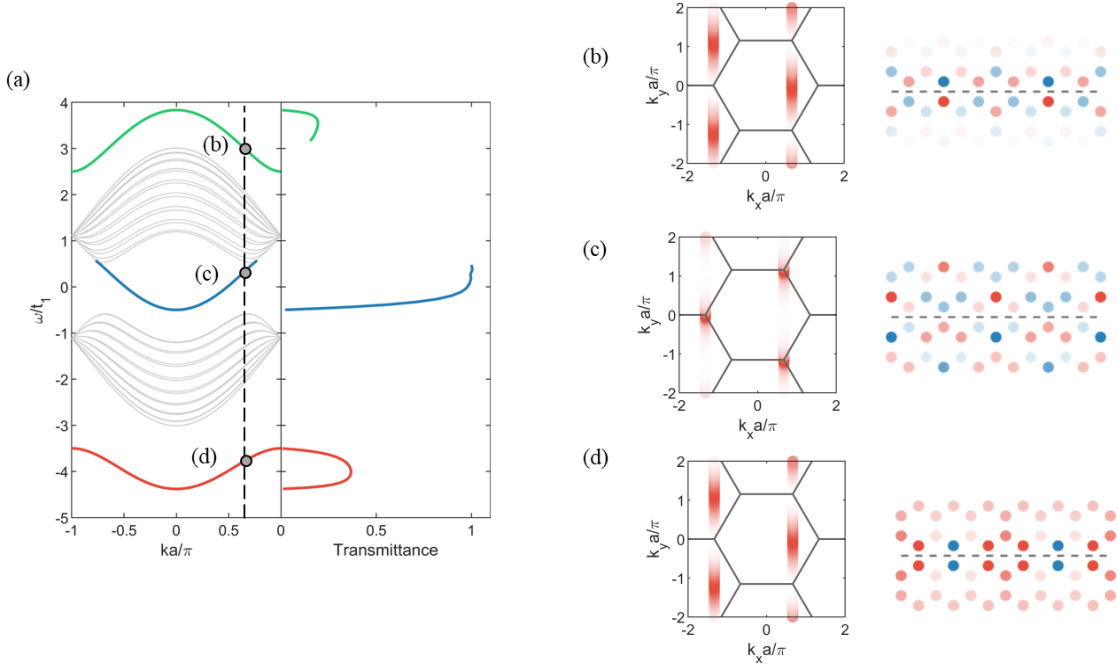

**Fig. S3.4.** (a) Band diagram (left panel) and transmittance (right panel) of edge states in a two-band tight-binding model with A and B sublattice. The tight-binding parameters are $t_0 = 3t_1$ and $\omega_0 = 0.5t_1$. (b), (c) ,(d) Modal profiles of edge states with wavenumber corresponding to the projected **K** point [marked in (a)] in *k*-space (left) and real space (right).

## S4. Valley Edge States, Bulk Bloch Modes, Bulk Valleys and Their Connections

VESs host momentum hotspots around high-symmetric **K/K'** points. In the main text, we argued that this specific momentum profile is rendered by the spectral neighborhood between the VESs and the **K/K'** valleys in the bulk band structure. The numerical results in Fig. 3(c) and (g)-(l) confirm this point. Below, we further solidify this argument by expanding the edge state with the bulk Bloch modes.

1. **Bloch Mode Expansion**

Consider that, at an interface along the *x*-direction between two 2D hexagonal photonic crystals (PhCs) with the same lattice constant [see Fig. 1(d) in the main text], edge states are supported, the electric fields of which are denoted by

$$\boldsymbol{E}_{e;j}(\boldsymbol{r}) = \boldsymbol{u}_j(\boldsymbol{r}) \exp(i\beta x),$$  Eq. S4.1

where the subscripts $j = 1,2$ indicate the lower and upper half-space crystals, respectively, and $\beta$ denotes the propagation wavenumber.

Since the bulk Bloch modes of a PhC constitute a complete basis, the edge state can be expressed by a linear superposition of the Bloch modes, whose electric fields are denoted by

$$\boldsymbol{E}_{b;j,n,\boldsymbol{k}_\beta}(\boldsymbol{r}) = \boldsymbol{v}_{j,n,\boldsymbol{k}_\beta}(\boldsymbol{r}) \exp(i\boldsymbol{k}_\beta \cdot \boldsymbol{r}),  \quad\text{Eq. S4.2}$$

where $n = 1,2,...$ label the photonic bands, $\boldsymbol{k}_\beta = \beta\hat{x} + k_y\hat{y}$, and $\boldsymbol{v}_{j,n,\boldsymbol{k}_\beta}$ is a periodic function with the same periodicity as the crystal. We then expand $\boldsymbol{E}_{e;j}$ with

$$\boldsymbol{E}_{e;j} = \sum_n \int dk_y\, \alpha_{j,n,\boldsymbol{k}_\beta} \boldsymbol{v}_{j,n,\boldsymbol{k}_\beta}(\boldsymbol{r}) \exp(i\boldsymbol{k}_\beta \cdot \boldsymbol{r}), \quad\text{Eq. S4.3}$$

where $\alpha_{j,n,\boldsymbol{k}_\beta}$ is the expansion coefficient of the Bloch mode, and $k_y$ traverses all possible values that make $\boldsymbol{k}_\beta$ (with fixed $\beta$) belong to the first Brillouin zone.

To derive the expansion coefficient $\alpha_{j,n,\boldsymbol{k}_\beta}$ that bridges the edge state and the bulk modes, we employ the Green's function technique. The Green's functions $\boldsymbol{G}_j(\boldsymbol{r};\boldsymbol{r}',\omega)$'s of two interfaced PhCs satisfy

$$\nabla \times \nabla \times \boldsymbol{G}_j(\boldsymbol{r};\boldsymbol{r}',\omega) - \omega^2 \varepsilon_j(\boldsymbol{r})\mu_0 \boldsymbol{G}_j(\boldsymbol{r};\boldsymbol{r}',\omega) = \boldsymbol{I}\delta(\boldsymbol{r}-\boldsymbol{r}'), \quad\text{Eq. S4.4}$$

where $\varepsilon_j$ denotes the permittivity profile of the crystal. $\boldsymbol{G}_j$ can be expanded with the Bloch modes

$$\boldsymbol{G}_j(\boldsymbol{r};\boldsymbol{r}',\omega) = \sum_n \int d\boldsymbol{k}\, \frac{\boldsymbol{E}_{b;j,n,\boldsymbol{k}}(\boldsymbol{r}') \otimes \boldsymbol{E}^*_{b;j,n,\boldsymbol{k}}(\boldsymbol{r})}{\omega_{j,n,\boldsymbol{k}}^2 - \omega^2}, \quad\text{Eq. S4.5}$$

where $\omega_{j,n,\boldsymbol{k}}$ is the real-valued eigenfrequency of the Bloch mode, and the $k$-space integration is performed over the first Brillouin zone. In Eq. S4.5, the Bloch modes are normalized by

$$\int d\boldsymbol{r}\, \varepsilon_j(\boldsymbol{r})\, \boldsymbol{E}_{b;j,n,\boldsymbol{k}_1}(\boldsymbol{r}) \cdot \boldsymbol{E}^*_{b;j,m,\boldsymbol{k}_2}(\boldsymbol{r}) = \delta_{nm}\delta(\boldsymbol{k}_1 - \boldsymbol{k}_2). \quad\text{Eq. S4.6}$$

Using the scattering wave formulation, the electric fields of the edge state in the lower (upper) PhC can be represented as the scattering fields induced by the perturbative polarizations in the upper (lower) PhC. Specifically, there are

$$\boldsymbol{E}_{e;1}(\boldsymbol{r}) = \omega^2\mu_0 \int_{y<0} d\boldsymbol{r}'\, \boldsymbol{G}_1(\boldsymbol{r};\boldsymbol{r}',\omega)\,(\varepsilon_2(\boldsymbol{r}') - \varepsilon_1(\boldsymbol{r}'))\boldsymbol{E}_{e;2}(\boldsymbol{r}'), \quad\text{Eq. S4.7(a)}$$

$$\boldsymbol{E}_{e;1}(\boldsymbol{r}) = \omega^2\mu_0 \int_{y>0} d\boldsymbol{r}'\, \boldsymbol{G}_2(\boldsymbol{r};\boldsymbol{r}',\omega)\,(\varepsilon_1(\boldsymbol{r}') - \varepsilon_2(\boldsymbol{r}'))\boldsymbol{E}_{e;1}(\boldsymbol{r}'). \quad\text{Eq. S4.7(b)}$$

Plugging Eqs. S4.2 and S4.5 into Eqs. S4.7, we obtain a set of linear equations concerning the modal expansion coefficients:

$$\alpha_{1,n,\mathbf{k}_\beta} = \frac{\omega^2}{\omega_{1,n,\mathbf{k}_\beta}^2 - \omega^2} \sum_m \int dk_y' \, h_{n,\mathbf{k}_\beta}^{m,\mathbf{k}_\beta'} \alpha_{2,m,\mathbf{k}_\beta'},  \qquad \text{Eq. S4.8(a)}$$

$$\alpha_{2,n,\mathbf{k}_\beta} = \frac{\omega^2}{\omega_{2,n,\mathbf{k}_\beta}^2 - \omega^2} \sum_m \int dk_y' \, g_{n,\mathbf{k}_\beta}^{m,\mathbf{k}_\beta'} \alpha_{1,m,\mathbf{k}_\beta'},  \qquad \text{Eq. S4.8(a)}$$

with

$$h_{n,\mathbf{k}_\beta}^{m,\mathbf{k}_\beta'} = \mu_0 \int dk_x \int_{y<0} d\mathbf{r} \, \mathbf{E}_{b;1,n,\mathbf{k}}^* \big(\varepsilon_2(\mathbf{r}) - \varepsilon_1(\mathbf{r})\big) \mathbf{E}_{b;2,m,\mathbf{k}_\beta'}(\mathbf{r}),  \qquad \text{Eq. S4.9(a)}$$

$$g_{n,\mathbf{k}_\beta}^{m,\mathbf{k}_\beta'} = \mu_0 \int dk_x \int_{y>0} d\mathbf{r} \, \mathbf{E}_{b;2,n,\mathbf{k}}^* \big(\varepsilon_1(\mathbf{r}) - \varepsilon_2(\mathbf{r})\big) \mathbf{E}_{b;1,m,\mathbf{k}_\beta'}(\mathbf{r}).  \qquad \text{Eq. S4.9(b)}$$

2. **Singe-band and k-p Approximations of Edge States and Discussions on Valley Edge States**

Equations S4.8 define an eigenvalue formulation for the edge states. Intuitively, they indicate that an edge state is mainly composed of its spectrally neighborhood Bloch modes due to the spectral dependence $1/(\omega_{1,n,\mathbf{k}_\beta} - \omega)$. To examine this intuition more concretely, we assume that the edge state is contributed from the Bloch modes of a single band, which exactly corresponds to the topological-trivial case studied in Fig. 3(c). Moreover, since we focus on the single band, the band label "$n$" shall be omitted hereafter. Further, we denote the wavenumber of a special bulk Bloch mode in the $j^{\text{th}}$ PhC, which, with $x$-component wavenumber $\beta$ as same as the edge state, is closest to the edge state in frequency, by

$$\mathbf{k}_{j,\beta}^N \equiv \beta \hat{x} + k_{j,y}^N \hat{y}.  \qquad \text{Eq. S4.10(a)}$$

The electric fields of this Bloch mode are accordingly denoted by

$$\mathbf{E}_{b;j,\mathbf{k}_{j,\beta}^N}(\mathbf{r}) = \mathbf{v}_{j,\mathbf{k}_{j,\beta}^N}(\mathbf{r}) \exp(i \mathbf{k}_{j,\beta}^N \cdot \mathbf{r}).  \qquad \text{Eq. S4.10(b)}$$

Employing the k-p approximation, the Bloch modes with wavenumber around $\mathbf{k}_{j,\beta}^N$ are approximated by

$$\mathbf{E}_{b;j,\mathbf{k}_\beta}(\mathbf{r}) \simeq \mathbf{v}_{j,\mathbf{k}_{j,\beta}^N}(\mathbf{r}) \exp(i \mathbf{k}_\beta \cdot \mathbf{r}).  \qquad \text{Eq. S4.10(c)}$$

With the notation settings and the k-p approximation summarized in Eqs. S4.10, we iteratively solve Eqs. S4.8. The initially trial solution of the edge state, denoted by $\mathbf{E}_{e;j}^{(0)}$, is chosen to be Bloch mode with wavenumber $\mathbf{k}_{j,\beta}^N$, i.e.,

$$\mathbf{E}_{e;j}^{(0)}(\mathbf{r}) = C_j \mathbf{E}_{b;j,\mathbf{k}_{j,\beta}^N}(\mathbf{r}),  \qquad \text{Eq. S4.11}$$

where $C_j$ is a constant. Then, we put Eq. S4.11 to the right-handed side of Eqs. S4.8 and update the solution. The 1$^{\text{st}}$ order solution of the edge state is then given by

$$\alpha_{1,\mathbf{k}_\beta}^{(1)} = C_2 \frac{\omega^2}{\omega_{1,\mathbf{k}_\beta}^2 - \omega^2} h_{\mathbf{k}_\beta}^{\mathbf{k}_{2,\beta}^N},  \qquad \text{Eq. S4.12(a)}$$

$$\alpha_{2,k_\beta}^{(1)} = C_1 \frac{\omega^2}{\omega_{2,k_\beta}^2 - \omega^2} g_{k_\beta}^{k_{1,\beta}^N}, \qquad \text{Eq. S4.12(b)}$$

with

$$h_{k_\beta}^{k_{2,\beta}^N} = \mu_0 \int dk_x \int_{y<0} d\boldsymbol{r}\, \boldsymbol{v}^*_{1,k_{1,\beta}^N}(\boldsymbol{r})(\varepsilon_2(\boldsymbol{r}) - \varepsilon_1(\boldsymbol{r}))\boldsymbol{v}_{2,k_{2,\beta}^N}(\boldsymbol{r}) \exp(i\boldsymbol{k}_{2,\beta}^N \cdot \boldsymbol{r} - i\boldsymbol{k} \cdot \boldsymbol{r}), \quad \text{Eq. S4.13(a)}$$

$$g_{k_\beta}^{k_{1,\beta}^N} = \mu_0 \int dk_x \int_{y>0} d\boldsymbol{r}\, \boldsymbol{v}^*_{2,k_{2,\beta}^N}(\boldsymbol{r})(\varepsilon_1(\boldsymbol{r}) - \varepsilon_2(\boldsymbol{r}))\boldsymbol{v}_{1,k_{2,\beta}^N}(\boldsymbol{r}) \exp(i\boldsymbol{k}_{1,\beta}^N \cdot \boldsymbol{r} - i\boldsymbol{k} \cdot \boldsymbol{r}). \quad \text{Eq. S4.13(b)}$$

Further, in Eqs. S4.12, we approximate the eigenfrequency of the Bloch modes $\omega_{j,k_\beta}$ by the Taylor expansion,

$$\omega_{j,k_\beta} \simeq \omega_{j,k_\beta^N} + m_{j\beta}^N (k_y - k_{j,y}^N)^2, \qquad \text{Eq. S4.14(a)}$$

where

$$m_{j\beta}^N \equiv \frac{1}{2} \frac{\partial^2 \omega_{j,k_\beta}}{\partial k_y^2}\bigg|_{k_y = k_y^N}. \qquad \text{Eq. S4.14(b)}$$

Plugging Eqs. S4.14 into Eqs. S4.12, and, then, evaluating Eqs. S4.3, we obtain that

$$\boldsymbol{E}_{e;1}(\boldsymbol{r}) \sim \boldsymbol{v}_{1,k_{1,\beta}^N}(\boldsymbol{r}) \exp(i\boldsymbol{k}_{1,\beta}^N \cdot \boldsymbol{r} - \alpha_1 y), \qquad \text{Eq. S4.15(a)}$$

$$\boldsymbol{E}_{e;2}(\boldsymbol{r}) \sim \boldsymbol{v}_{2,k_{2,\beta}^N}(\boldsymbol{r}) \exp(i\boldsymbol{k}_{2,\beta}^N \cdot \boldsymbol{r} + \alpha_2 y), \qquad \text{Eq. S4.15(b)}$$

where the decaying wavenumber $\alpha_j$ is given by

$$\alpha_j = \sqrt{\left|\frac{\omega_{j,k_\beta^N} - \omega}{m_{j\beta}^N}\right|}. \qquad \text{Eq. S4.15(c)}$$

Equations S4.15 are derived by extending the integration range of $k_y$ from $-\infty$ to $\infty$, and picking up the pole contribution using Cauchy's residue theorem. We can continue the iteration process by plugging Eqs. S4.15(a) and S4.15(b) into the right-handed side of Eqs. S4.8, and repeat the above steps. The similar analysis shows that Eqs. S4.15 consistently hold under the same approximations. Moreover, the decaying wavenumber of the VESs, according to Eq. S4.15(c), scale with $\sqrt{|\omega_{j,k_\beta^N} - \omega|}$. To verify this deduction, we consider the single-band tight-binding model that has been studied in Fig. 3(c). To tune the decaying wavenumber, the interface hooping parameter is varied from $t_0 = t_2$ to $t_0 = 4t_2$. The decaying wavenumber of the edge state with wavenumber corresponding to the projected **K** point as a function of $t_0$ is plotted in Fig. S4.1. The exact results and the approximation results obtained with Eq. S4.15(c) agree well with each other. Note that, to evaluate the decaying wavenumber, the exact frequency of the edge state is used.

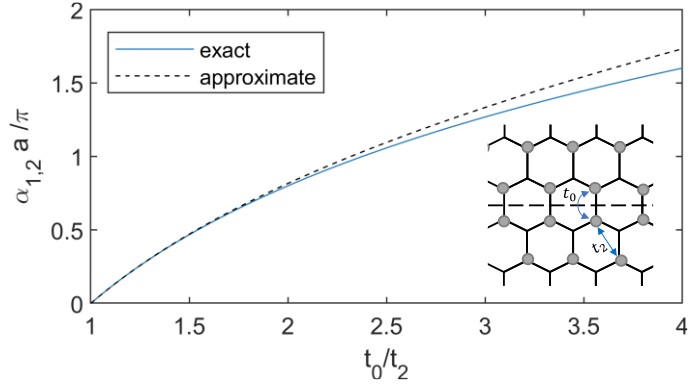

**Fig. S4.1. Numerical validation of decaying wavenumber predicted by Eq. S4.15(c) (approximate).** The single-band tight-binding model as Fig. 3(c) in the main text is considered, and the interface hooping parameter $t_0$ is varied from $t_0 = t_2$ to $t_0 = 4t_2$, and the edge state with wavenumber corresponding to the projected **K** point in the upper branch of the edge states [cf. Fig. 3(c)] is chosen.

When multiple bands are taken into consideration, the above discussions are still useful. For instance, consider the two-band tight-binding model with A and B sublattices. The edge states, whose wavenumbers correspond to the projected **K/K'** points and lie between the two bands, have their spectrally closet Bloch modes at the **K/K'** valleys in both the upper and lower bands. Similar to the single-band case, the edge states can be expressed as the linear combination of the Bloch modes centralized around the **K/K'** valleys of the two bands, and, thus, have modal profiles close to the Bloch modes of the **K/K'** points. Alternatively, if the edge state is much closer to one band than the others, the single-band approximation can be adopted.

## S5. Edges states generated on armchair interface

The PhCs discussed previously are all terminated and interfaced along zig-zag interface. Here, we demonstrate a type of edge states generated at an armchair interface, illustrate their momentum profile and interpret the reason why they do not support bending immune transport.

The PhCs in Fig. S5(a) on either side are cut along the armchair direction, and are glided to maintain the overall hexagonal crystal structure. We set the interface direction to be along the *x* axis. The reciprocal lattices are demonstrated in Fig. S5(b). If we consider the projection of the bulk Bloch modes onto the direction of the armchair interface, we see that the opposite valleys **K** and **K'** mix (gray dotted lines in Fig. S5(b)). Figure S5(c) demonstrates the projected band diagram, with blue and red dots representing bulk and edge states, respectively. We then focus on the edge state with normalized wavenumber equals to 0 or $\frac{\pi}{2a}$, which corresponds to the cutlines through K and K' in Fig. S5(b). As shown in Fig. S5(d), the momentum profile of the edge state in this case is not only localized on **K** points, but also on **K'** points. Such momentum profile does not coincide with our definition of valley edge states. Comparing the momentum profiles of incident, transmitted and reflected modes, we can see that there is no superiority for $\psi_i, \psi_t$ overlap compared with $\psi_i, \psi_r$ overlap and the bending immunity cannot be realized.

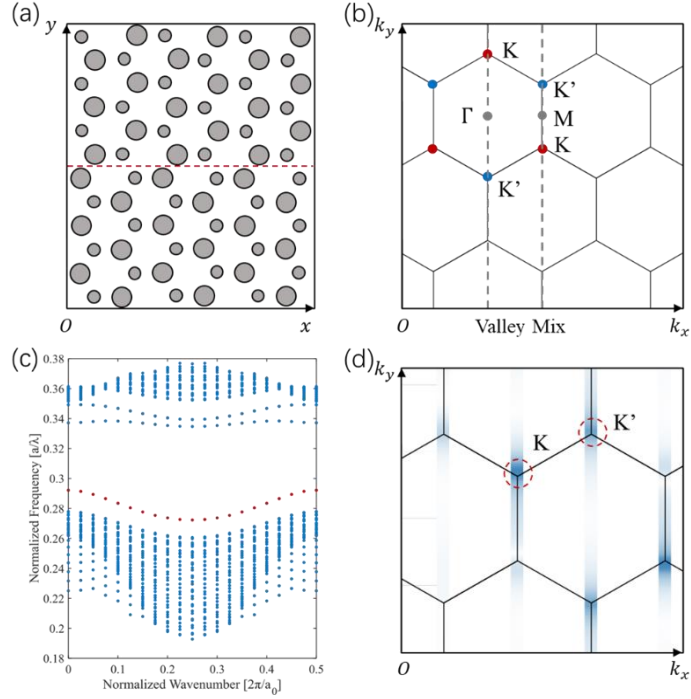

**Fig. S5. Illustration of the edge states along the armchair interface.** (a) An example of PhC with armchair interface. (b) The reciprocal lattice of the PhC, of which the armchair interface is set along the *x*-direction. (c) Projected band diagram of bulk modes (blue) and edge states (red). (d) Momentum profile of the edge state of which the wavenumber corresponds to the projection of **K** and **K'** points.

## S6. Topological photonics based on 2D Zak Phase

Besides valley topology, there are other time-reversal-symmetry preserved topological systems. Here we consider a topological invariant named 2D Zak phase. Zak phase[3] was initially studied in 1D system based on Su-Schrieffer-Heeger (SSH) modal[4] and then expanded to 2D systems. 2D Zak phase intimately relates to 2D bulk polarization and characterizes the delocalization of Wannier center[5], which can be calculated by

$$P_{Zak} = \frac{1}{2\pi}\int dk_1 \int dk_2 \, \text{Tr}[\mathcal{A}(k_1\boldsymbol{b}_1 + k_2\boldsymbol{b}_2)] \qquad Eq.\,S6.1$$

where $k_1 \in [0\ 1]$ and $k_2 \in [0\ 1]$ are the normalized wavenumbers, $\boldsymbol{b}_1, \boldsymbol{b}_2$ are reciprocal lattice vectors, and $\mathcal{A}$ is the Berry connection. When the Wannier center is localized at the lattice point, or the center of PhC lattice, the 2D Zak phase is $P_{Zak} = (p_{k_1}, p_{k_2}) = (0,0)$. When the Wannier center is delocalized from the lattice point, or the center of PhC lattice, the 2D Zak phase becomes nonzero. For lattices belong to dihedral group, the nonzero 2D Zak phase, or the so-called bulk polarization can only obtain discretized value, and the topological invariant can thus be well defined[5]. For square (Fig. S6.1(a)) and rhombic (Fig. S6.1(b)) lattices, $P_{Zak}$ can be $(0,0)$ (trivial) or $(1/2, 1/2)$ (non-trivial). Notably, the distribution of the eigenmodes in the real space (right panels of Fig. S6.1(a), (b)) indicates the dislocation of Wannier centers.

Interfacing PhCs with trivial and non-trivial 2D Zak phases can generate topological edge states. Different from the gapless edge states reported in the topological Chern systems, this type of edge states exists within the bandgap. Topological PhCs based on 2D Zak phase have been widely reported for supporting high order corner states. But as for 1D propagating edge states, 2D Zak phase itself does not provide protective mechanism for their robust propagation.

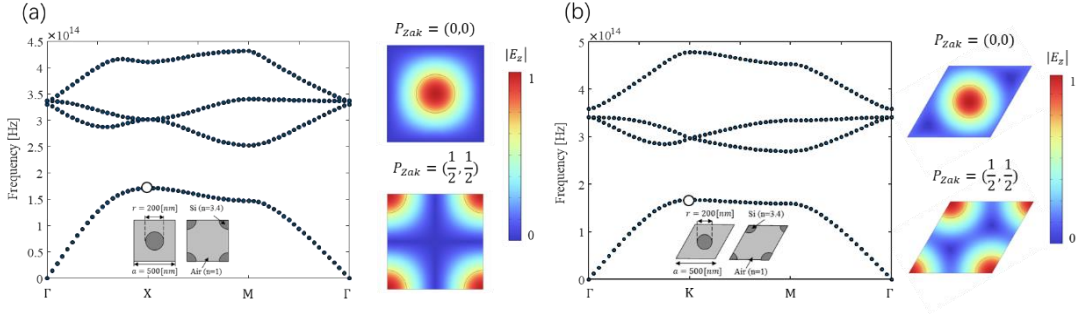

**Fig. S6.1. Illustration of PhCs with trivial and non-trivial 2D Zak phase.** (a) shows a square lattice PhC and (b) shows a triangular lattice PhC. Both figures include band diagrams and modal profiles at high symmetry points (**X** for (a) and **K** for (b)). For PhCs with trivial and non-trivial 2D Zak phases, the modal profiles have hotspots on different positions and exhibit different parities.

Figure S6.2(a) demonstrates the field profile of an edge state propagating through a 90° bending PhC waveguide formed by square lattice considered in Fig. S6.1(a). The 2D Zak phase for either half of PhC of the 1st band is different, (0,0) and (1/2,1/2), respectively. The simulation results illustrate strong reflection due to the bending. The reflection is also evident in the transmittance spectrum (Fig. S6.2(b)). The overall poor transmittance confirms that 2D Zak phase itself does not naturally support robustness against sharp bending.

However, things in a 120° bending waveguide formed by triangular lattice PhCs (demonstrated in Fig. S6.1(b)) with contrasting (0,0) and (1/2,1/2) 2D Zak phases (in the 1st band). change Figure S6.2(c) shows no significant reflections for an edge state propagating in this bending waveguide (the wavevector of the edge state is set to be equal to the projected **K** point). This observation is further confirmed in the transmittance spectrum (Fig. S6.2(d)), which demonstrates a flat ~100% passband. Such high-transmittance bending properties of the edge states cannot be attributed to the topological protection of 2D Zak phase. Instead, the modal profile coinciding with valley edge states is responsible for such bending immune properties (see Fig. 4 in the main text for the momentum profiles of the edge state).

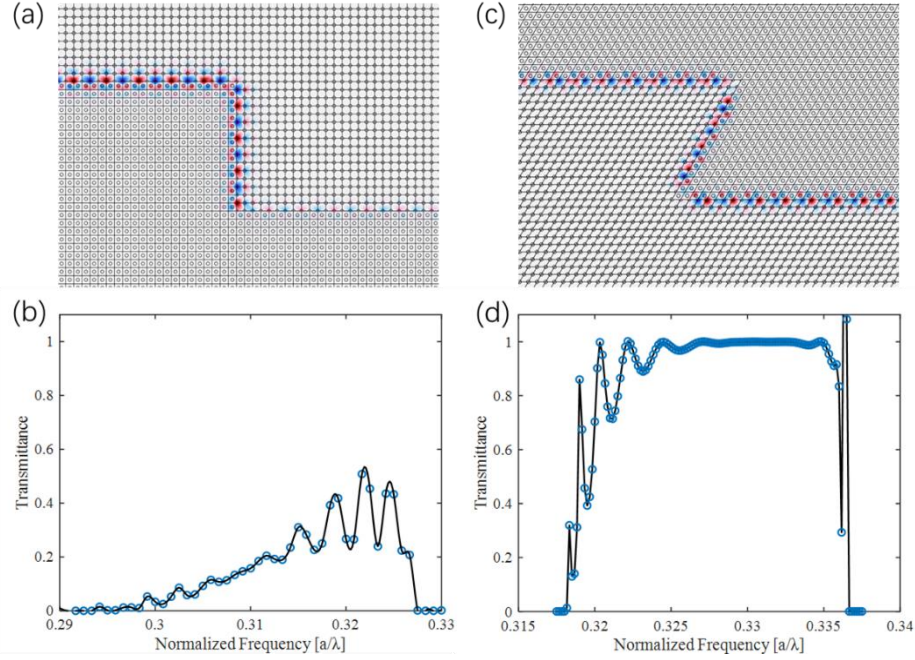

**Fig. S6.2. The bending properties for topological edge states based on 2D Zak phase.** (a) and (b) show that 2D Zak phase itself cannot support bending immune transmission in square lattice while (c) and (d) show that when the edge states are implemented in the triangular lattice and they become valley edge states (see Fig. 4 in in the main text for the momentum profile), they gain the potential of bending immunity.

## S7. Quasi-valley edge states in rhombic lattice

The concept of valley edge states and its interpretation of bending immune transmittance can be expanded to PhCs based on lattice other than triangular or hexagonal lattice (without $C_6$ or $C_3$ symmetry). Based on this understanding, we utilize rhombic lattice PhC to construct quasi-valley edge states along the "zig-zag" interface (red dotted line in Fig. S7.1(a)). The projected band diagram of this PhC onto the direction of the zig-zag interface is illustrated in Fig. S7.1(b), where the blue dots represent the bulk modes while the red dots represent the edge states. If we set the wavenumber according to the projection of $\mathbf{O}, \mathbf{M}, \mathbf{\bar{N}}$ points, which are a group of specific reciprocal lattice points similar to **K** points in triangular or hexagonal lattice, we can get a modal profile similar to the valley edge states. As illustrated in Fig. S7.1(c), the momentum profile shows hotspots near the $\mathbf{O}, \mathbf{M}, \mathbf{\bar{N}}$ points. Based on this property, such edge states are named as quasi-valley edge states

Though no longer possessing $C_3$ or $C_6$ symmetry, the momentum matching mechanism still works. As shown in Fig. S7.2, the hotspots of the incident and transmitted modes coincident with each other (both around $\mathbf{O}, \mathbf{M}, \mathbf{\bar{N}}$) while hotspots of the reflected modes distribute on separate group of reciprocal lattice points ($\mathbf{\bar{O}}, \mathbf{\bar{M}}, \mathbf{N}$). Although as $(k_x, k_y)$ moves away from the first Brillouin zone, the hotspots will gradually deviate from the $\mathbf{O}, \mathbf{M}, \mathbf{\bar{N}}$ or $\mathbf{\bar{O}}, \mathbf{\bar{M}}, \mathbf{N}$ points (red dotted circle in Fig. S7.1(c)) and the coincidence between the incident and transmitted modes thereby degenerates. Nevertheless, the high contrast between the overlap term $\int d\mathbf{k}_i \int d\mathbf{k}_t \psi_t^*(\mathbf{k}_t)\psi_i(\mathbf{k}_i)\Delta\varepsilon(\mathbf{k}_t - \mathbf{k}_i)$ and $\int d\boldsymbol{k}_i \int d\boldsymbol{k}_r \psi_r^*(\boldsymbol{k}_r)\psi_i(\boldsymbol{k}_i)\Delta\varepsilon(\boldsymbol{k}_r - \boldsymbol{k}_i)$, together with the fact that the main components are distributed in the first Brillouin zone, make the edge states still maintain the tolerance to sharp bending.

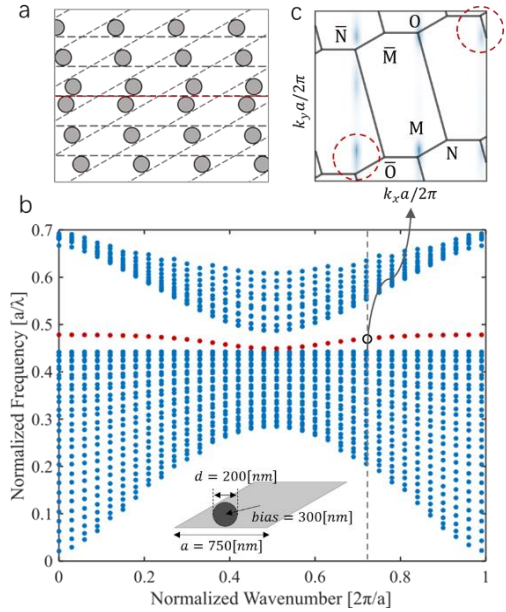

**Fig. S7.1.** (a) The structure of rhombic lattice PhC that supports quasi-valley edge states. (b) The band diagram of such PhC including bulk modes (blue) and edge states (red). (c) The momentum profile of quasi-valley edge states of which the wavenumber corresponding to the projection of $\mathbf{O}, \mathbf{M}, \overline{\mathbf{N}}$ points.

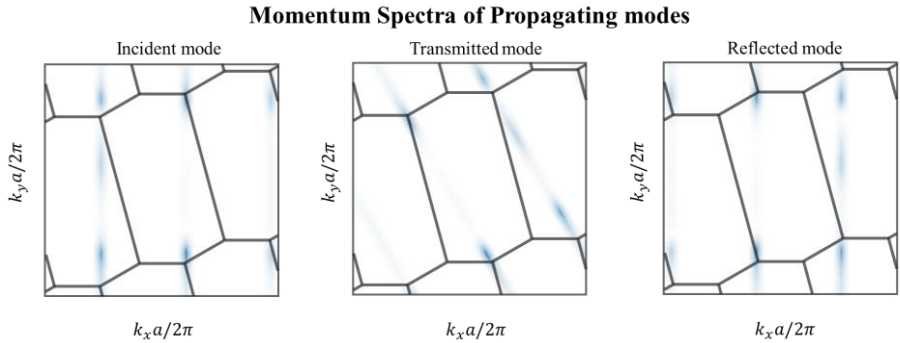

**Fig. S7.2. The momentum profiles of incident (left), transmitted (middle) and reflected (right) modes.**

Based on this understanding, high-transmittance bending waveguides with various bending angle can be constructed utilizing rhombic lattice PhCs. Figure S7.3 demonstrates bending waveguides with $53.1°, 36.8°, 30°$ acute angle. The side length of the lattice is 750 [nm], the radius of the Si rods is 100 [nm]. The field distributions for the edge states propagating through Z-shaped waveguides are illustrated in (a)-(c) and the transmittance spectra are shown in (d). Both of them show high-transmittance passband within the bandwidth of the edge states except a few dips due to the excitations of the resonant corner states (to be discussed later).

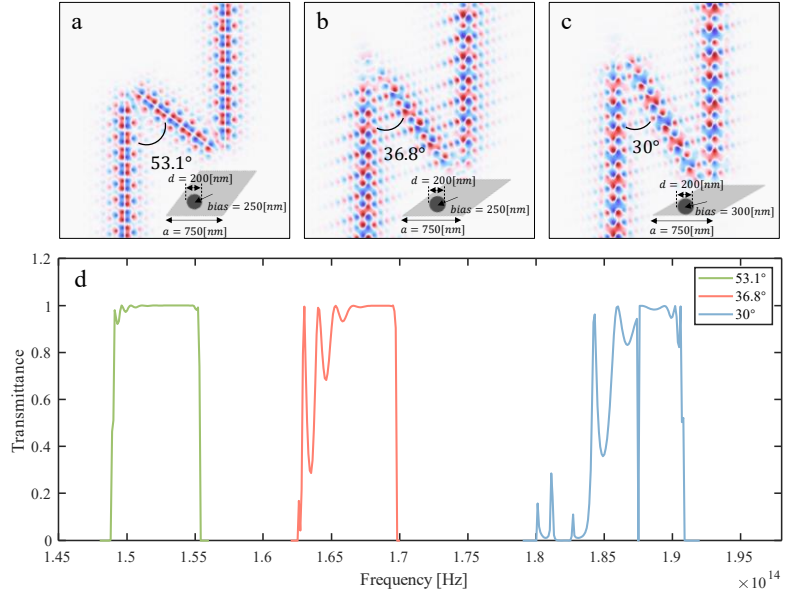

**Fig. S7.3. The transmission properties for bending waveguides with different bending angle (acute angles are 53.1°, 36.8°, 30°).** All of them support high transmittance passband.

We want to emphasize that other additional conditions should be taken into account to guarantee the realization of bending immune waveguides. For instance, it is possible that at the operating frequency $f_0$, there possibly exist many propagating modes, such as $\psi_i(k_1)$, $\psi_i(k_2)$, $\psi_i(k_3)$, $\psi_i(k_4)$ shown in Fig. S7.4(b). In this case, the complex coupling between them will increase the possibility of reflection coupling and result in low bending transmittance. There is another interesting example corresponding to the dip in the 30° bending transmittance spectrum (Fig. S7.4(d) and Fig. S7.4(c)). As illustrated in Fig. S7.4(d), this dip is caused by a resonance existing at the bending corner. According to our calculation, we deduce that it might be a corner state. We will not discuss the origin of the corner state here, for which the associated discussions can be found in Refs. [6-9], instead, we want to point out that the participation of resonances can damage the bending immune propagation.

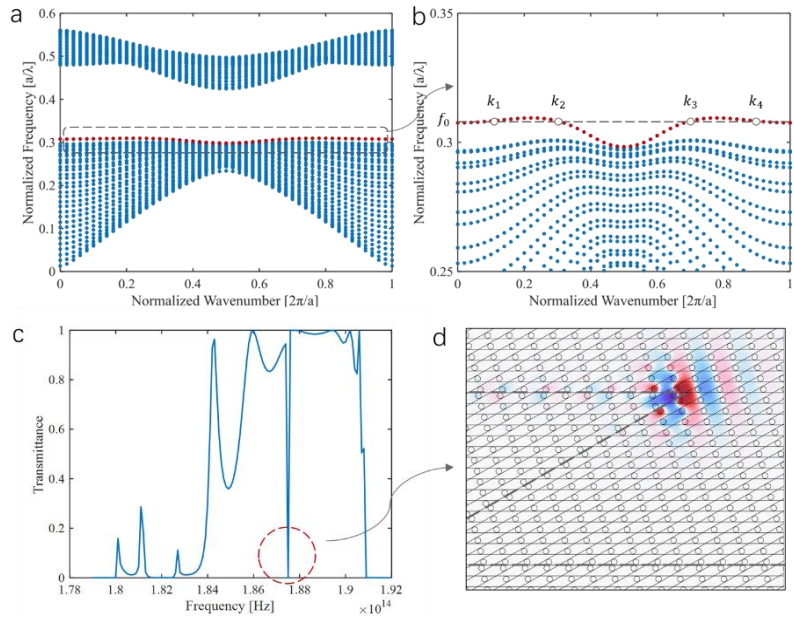

**Fig. S7.4. Two examples to show that two additional factors should be paid attention for bending immune waveguides.** (a) and (b) At one operating frequency, there are possibly many edge modes and their complicated interactions may result in degradation of bending transmittance. (c) and (d) When a resonance is formed at a corner, the mode coupling between propagating modes and corner resonance can also damage the bending immune properties.

## S8. Simulation settings for Fig. 4 in the main text

In Fig. 4, we illustrate several examples of bending immune edge states. Here, the simulation settings are summarized in Fig.S8.1 and Table S8.1.

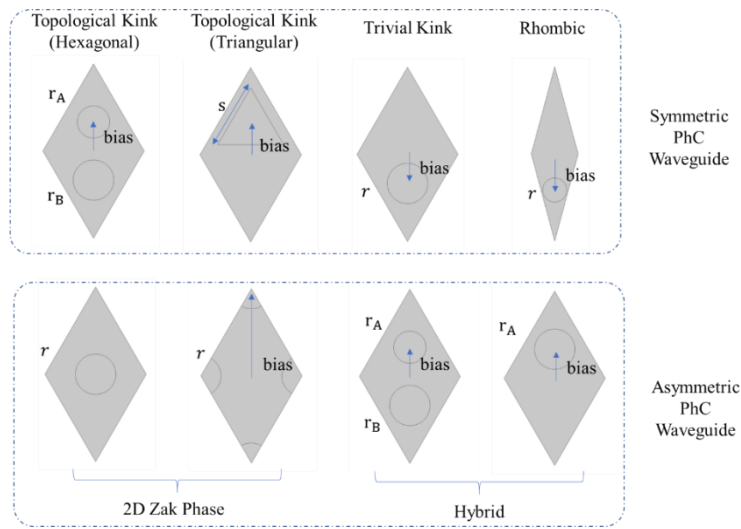

**Fig. S8.1. The structure of the lattices in Fig. 4 of the main text.**

| PhC Waveguide | Lattice constant | Size of Si Structure | Bias | Simulation Frequency |
|---|---|---|---|---|
| **Topological Kink (Hexagonal)** | 500[nm] | $r_A = 160$[nm], $r_B = 200$[nm] | $\dfrac{500[nm]}{2\sqrt{3}}$ | 172.29 [THz] |
| **Topological Kink (Triangular)** | 500[nm] | s = 325[nm] | $\dfrac{175[nm]}{2\sqrt{3}}$ | 165.32 [THz] |
| **Trivial Kink** | 500[nm] | r = 200[nm] | $\dfrac{500[nm]}{2\sqrt{3}}$ | 178.69 [THz] |
| **Rhombic** | 750[nm] | r = 200[nm] | 300[nm] | 188.9 [THz] |
| **2D Zak Phase** | 500[nm] | r = 200[nm] | $\dfrac{500\sqrt{3}[nm]}{2}$ | 201.3 [THz] |
| **Hybrid** | 500[nm] | $r_A = 160$[nm], $r_B = 200$[nm] | $\dfrac{500[nm]}{2\sqrt{3}}$ | 174.82 [THz] |

**Table. S8.1. Simulation parameters corresponding to Fig. 4 of the main text.**